\documentclass[fleqn,usenatbib]{mnras}

\usepackage[T1]{fontenc}
\usepackage{ae,aecompl}

\usepackage{graphicx}	
\usepackage{amsmath}	
\usepackage{amssymb}	
\defcitealias{prime1}{Primeval~I}
\defcitealias{prime2}{Primeval~II}
\defcitealias{prime3}{Primeval~III}
\defcitealias{prime4}{Primeval~IV}
\defcitealias{prime5}{Primeval~V}
\defcitealias{prime6}{Primeval~VI}
\defcitealias{burg06}{Burgasser et al. 2006}
\defcitealias{wool05}{Woolf et al. 2005}

\title[The first wide M + L extreme subdwarf binary]{Primeval very low-mass stars and brown dwarfs -- VII. The discovery of the first wide M + L extreme subdwarf binary
}
\author[Z. H. Zhang]{ZengHua Zhang$^{1,2}$\thanks{E-mail:
zenghuazhang@hotmail.com}\\
$^{1}$School of Astronomy and Space Science, Key Laboratory of Ministry of Education, Nanjing University, Nanjing 210023, China \\
$^{2}$GEPI, Observatoire de Paris, Universit{\'e} PSL, CNRS, 5 Place Jules Janssen, F-92190 Meudon, France 
}

\date{Accepted 2019 August 2. Received 2019 July 23; in original form 2019 June 10}

\pubyear{2019}

\begin{document}
\label{firstpage}
\pagerange{\pageref{firstpage}--\pageref{lastpage}}
\maketitle

\begin{abstract}
I present the discovery of the first wide M + L extreme subdwarf binary system Gaia J0452$-$36AB. The binary is located at a distance of 137.27$^{+0.68}_{-0.67}$ pc with a projected separation of 15828$\pm$78 au. I classified Gaia J0452$-$36AB as esdM1 and esdL0 subdwarfs, respectively. Gaia J0452$-$36AB have typical halo kinematics, metallicity of [Fe/H] $\approx -1.4$, and temperature of $\sim$ 3550 and 2600 K, respectively. Gaia J0452$-$36AB is a pair of very low-mass stars with masses of 0.151$^{+0.029}_{-0.019}$ and 0.0855$^{+0.0014}_{-0.0010}$ M$_{\sun}$, and is a gravitationally bound system. I tested the metallicity consistency of existing M subdwarf classification schemes with Gaia J0452$-$36AB and a sample of M and L subdwarfs with known metallicity. I found that the metallicity of each M subclass defined by the the metallicity index $\zeta_{\rm CaH/TiO}$ is not consistent from mid-to-late M subtypes. Because late-type M and L subdwarfs have dusty atmospheres and high surface gravity which have significant impacts on CaH and TiO indices that used in the classification. The metallicity scale of late-type M subdwarfs would be overestimated by the $\zeta_{\rm CaH/TiO}$ index. I discussed the mass range of M subdwarfs, and explained the lack of late-type M extreme and ultra subdwarfs, and decreasing binary fraction from sdM, to esdM, and usdM subclasses. The four M subclasses have different mass ranges. The comparison between M subclasses is between populations in different mass ranges. I also present the discovery of Ruiz 440-469B, an M8 dwarf wide companion to a cool DA white dwarf, Ruiz 440-469. 
\end{abstract}




\section{Introduction}
Field red dwarfs have temperature of $\sim$ 2000--4000 K, and have molecules (TiO, CaH, and H$_2$O) in their atmospheres. They are very low-mass stars with mass of $\sim$ 0.08--0.6 M$_{\sun}$. Red subdwarfs have subsolar metallicity and lower opacity than red dwarfs. Consequently, red subdwarfs have bluer colours than red dwarfs and appear below the main sequence on the Hertzsprung--Russell diagram \citep[HRD;][]{hert09,russ14}. Red subdwarfs have higher temperature ($T_{\rm eff}$) than red dwarfs of the equivalent mass \citep[fig. 9 in ][hereafter \citetalias{prime2}]{prime2}. 
Molecular features start to appear in the spectra of red subdwarfs at lower temperature than red dwarfs \citep[e.g. fig. 9;][]{jao08}. Red subdwarfs have smaller radii than red dwarfs of the equivalent $T_{\rm eff}$ \citep{kess19}. Red subdwarfs also have lower and narrower mass range than red dwarfs. 

In general, red dwarfs are orbiting in the Galactic disc, while older red subdwarfs are kinematically associated with the halo or thick disc with relatively higher space velocities. Red subdwarfs have spectral types of late-type K, M, and early-type L. Early-type M subdwarfs have lower $T_{\rm eff}$ than M dwarfs with equivalent subtypes due to much lower masses. Mid- and late-type M, and L subdwarfs have similar or higher mass than dwarfs with equivalent subtypes thus also have higher $T_{\rm eff}$ \citep[fig. 4;][hereafter \citetalias{prime3}]{prime3}.

Dwarfs with spectral types of $\geq$ M7 have dust formation in their atmospheres \citep{jone97} and thus are referred to as ultracool dwarfs \citep[UCD; e.g.][]{kirk97}. Likewise, subdwarfs with spectral types of late-type M or L are called ultracool subdwarfs \citep[UCSD; e.g.][]{lepi03a}. UCDs and UCSDs are extremely low-mass stars or brown dwarfs with mass of $\la$ 0.1 M$_{\sun}$.  

The Kapteyn's star is the nearest cool subdwarf \citep[at 3.93 pc;][]{gaia18} with a spectral type of sdM1 and a mass of $\sim$0.2 M$_{\sun}$ \citepalias[fig. 9;][]{prime2}. It was first discovered as a high proper motion star (\citealt{kapt97}; R. Innes; \citealt{gill99}). Then it was classified as an M subdwarf \citep{kuip40} after the discovery of the HRD and definition of cool subdwarfs \citep{kuip39}. 
To date, thousands of M subdwarfs \citep{zha13,savc14,zhon15,jao17,zhan19}, hundreds of late-type M subdwarfs \citep{lepi03,burg07,lepi08,lodi12,lodi17,kirk16}, and about 66 L subdwarfs (\citealt{burg03}; \citealt{kirk14}; \citealt[hereafter \citetalias{prime1}]{prime1}; \citetalias{prime3}; \citealt[hereafter \citetalias{prime4}]{prime4}) have been discovered with modern sky surveys. About 41 T subdwarfs have also been discovered \citep[e.g.,][]{burg02,pinf12,mace13,burn14} and studied in the literature \citep[hereafter \citetalias{prime6}]{prime6}. 

\citet{gizi97} classified M dwarfs/subdwarfs into three metallicity subclasses: M dwarfs (M V), M subdwarf (sdM), and extreme subdwarf (esdM) based on their broad band absorption features (CaH and TiO). The metallicity consistency of these subclasses are tested with observed HRD of globular clusters for early-type M subdwarfs. Then, \citet{lepi07}  defined a metallicity index ($\zeta_{\rm CaH/TiO}$) with CaH2, CaH3, and TiO5 spectral indices \citep{reid95}, and revised the classification into four subclasses: M dwarfs (dM), sdM, esdM, and ultra subdwarf (usdM). The $\zeta_{\rm CaH/TiO}$ index was slightly refined a few times \citep{dhit12,lepi13,zhan19}. The metallicity consistency of M subclasses are tested with wide binaries of early- and mid-type M subdwarfs. However, the test was not possible across late-type M subdwarfs due to the lack of wide UCSD binaries. 

There are about four binary systems with late-type M or L subdwarf components known in the literature. LSR 1610$-$0040 \citep{lepi03} is an unresolved metal-poor ([Fe/H] $\approx -$1.0) binary \citep{kore16} that is composed of a late-type M subdwarf and a degenerate brown dwarf \citepalias[D-BD;][]{prime6}. HD 114762AB is a close sdF9+sdM9 binary with [Fe/H] $\approx -$0.7 \citep{bowl09}. SDSS J1416+13AB is a wide sdL7+sdT7.5 subdwarf binary with Fe/H] $\approx -$0.3 \citep{burn10}. GJ 660.1AB is a wide sdM1+sdM7 binary with [Fe/H] $\approx -$0.63 \citep{agna16}.

Wide F/G/K/M + L subdwarf binaries with well-constrained properties are ideal benchmarks to: (1) test metallicity consistency in the classification of M subdwarfs \citep[e.g.,][]{lepi07}; (2) calibrate precise metallicity measurements of M and L subdwarfs in the near infrared  \citep[NIR; e.g.,][]{roja12,newt15}; and (3) test the performance of ultracool atmospheric models and very low-mass evolutionary models in the subsolar metallicity domain (\citealt{alla95,alla14,chab97,bara97,burr01}; Marley et al. 2019, in prep.). However, such binaries have not been discovered in the literature. Therefore, I conducted a search for wide L subdwarf binaries with the second data release (DR2) of {\sl Gaia} \citep{gaia18} and the Visible and Infrared Survey Telescope for Astronomy's (VISTA) Hemisphere Survey \citep[VHS;][]{mcma13}. 

This is the seventh paper of a series titled {\sl Primeval very low-mass stars and brown dwarfs}. The first to the fifth papers of the series are focused on L subdwarfs, transitional brown dwarfs (T-BDs), and the substellar transition zone (\citetalias{prime1}; \citetalias{prime2}; \citetalias{prime3}; \citetalias{prime4}; \citealt[hereafter \citetalias{prime5}]{prime5}). Population properties of metal-poor D-BDs are discussed in the sixth paper \citepalias{prime6}. Some of these results are summarised in \citet{zha18c}. In this paper, I present the first wide M + L extreme subdwarf binary discovered with the {\sl Gaia} and VHS surveys. Candidate selection is described in Section \ref{ssel}. Sections \ref{sobs} and \ref{scha} present spectroscopic observations and characterization of two wide binary systems, respectively. Metallicity consistency and mass ranges of M subclasses are discussed in Sections \ref{smc} and \ref{sms}, respectively. Section \ref{scon} presents the summary and conclusions.

\begin{figure*}
\begin{center}
   \includegraphics[width=\textwidth]{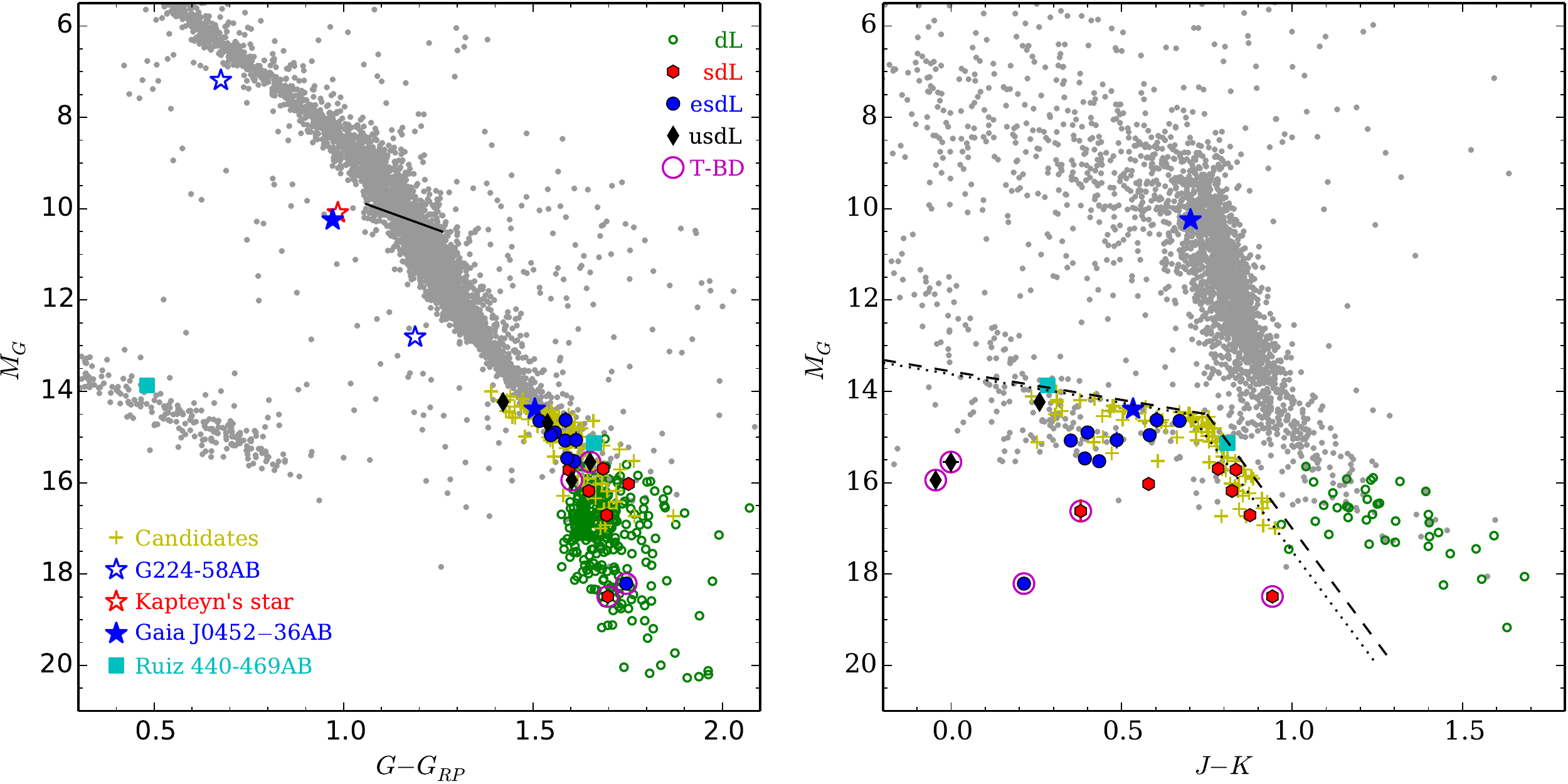}
\caption[]{HRDs for Gaia J0452$-$36AB (blue five-pointed stars), Ruiz 440-469AB (cyan squares), and other L subdwarf candidates (yellow crosses) selected from {\sl Gaia} and VHS in comparison to the Kapteyn's star (sdM1; red open five-pointed star), G224-58AB \citep[esdK5+esdM5.5; blue open five-pointed stars;][]{zha13,pavl15}, L subdwarfs \citepalias[fig. 21,][]{prime4}, and field objects. The $J-Ks_{\rm V}$ colour of Gaia J0452$-$36B, Ruiz 440-469B, and L subdwarf candidates are transformed to $J-K$ colour in the right-hand panel by $K - Ks = -0.05$ \citepalias[fig. 7,][]{prime6}. Metal-poor T-BDs are indicated with magenta circles. Grey dots are 5000 objects selected from Gaia DR2 and Large Area Survey of UKIDSS with distance < 100 pc, $180\degr < RA < 220\degr$ and $0\degr < Dec. < 20\degr$. The two grey sequences are white dwarfs (left) and main-sequence stars (right). Some bright field stars are scattered in the right-hand panel mostly because they are too bright in the UKIDSS fields. Dashed and dotted lines in the right hand panel indicate my selection cuts for L subdwarf candidates in UKIDSS ($J-K$) and VISTA ($J-Ks$), respectively. A black line on the left hand panel [from (1.06, 9.9) to (1.26, 10.5)] indicates a gap among M dwarfs \citep{jao18} associated with mixing of $^3$He that reduced the nuclear fusion rate around 0.34--0.36 M$_{\odot}$ \citep{bara18,macd18}.}
\label{hrd}
\end{center}
\end{figure*}

\begin{figure*}
\begin{center}
   \includegraphics[width=\textwidth]{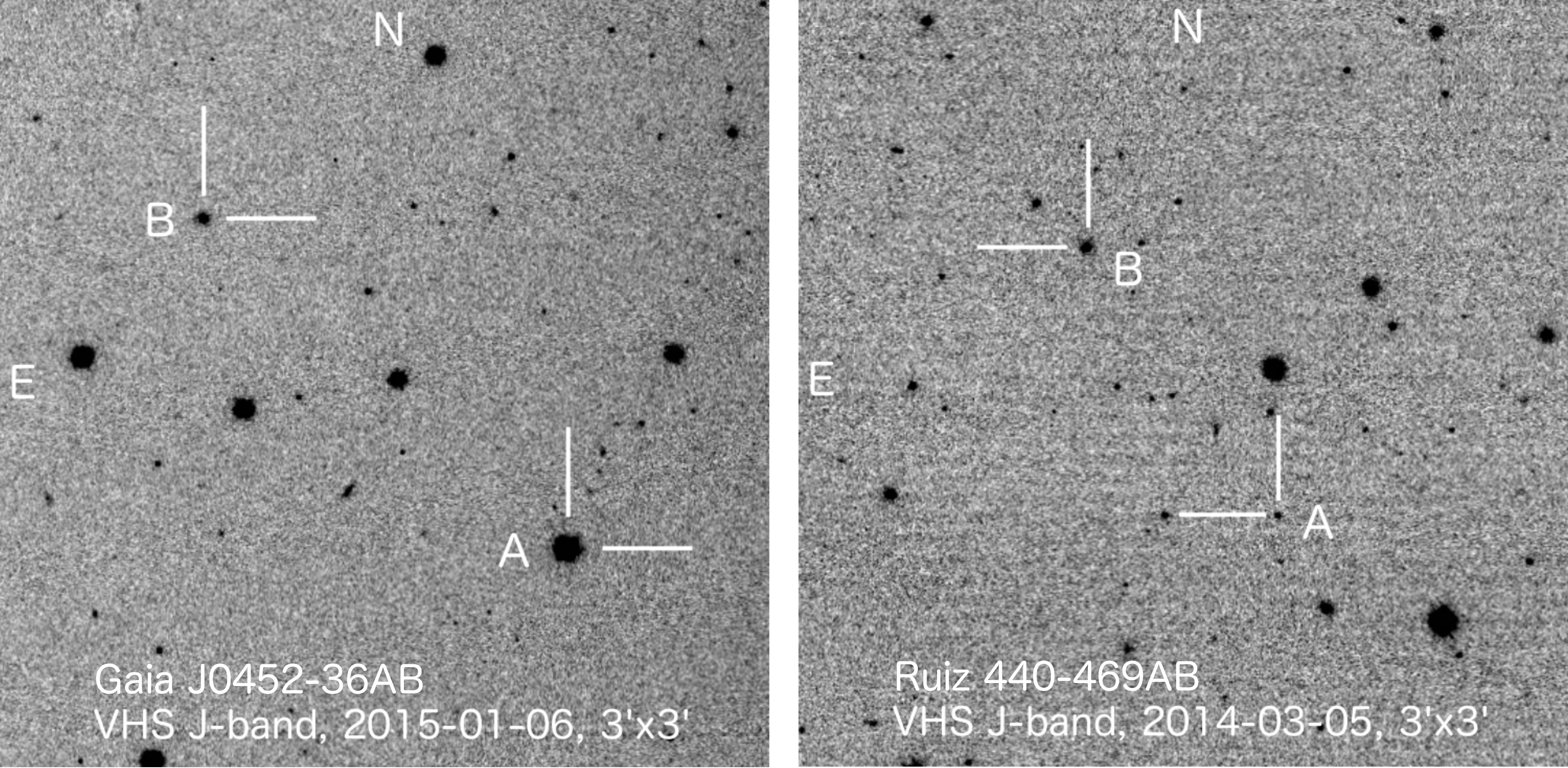}
\caption[]{VHS $J$-band images of fields around Gaia J0452$-$36AB and Ruiz 440-469AB, which are indicated with white bars (20 arcsec in length). Both fields are 3 arcmin on a side with north up and east to the left. Gaia J0452$-$36AB are moving toward southeast (138\fdg65) with a proper motion of 224.7 mas yr$^{-1}$. Ruiz 440-469AB are moving toward southwest (258\fdg56) with a proper motion of 210.9 mas yr$^{-1}$}
\label{vhsj}
\end{center}
\end{figure*}

\section{Candidate selection}
\label{ssel}
\subsection{L subdwarf candidates}
L subdwarfs have subsolar metallicity and are members of the Galactic thick disc or halo. Therefore, they have distinctive photometric colours, proper motion, and tangential velocity from field objects \citepalias{prime1}.  I conducted a search for L subdwarf candidates by a combined used of {\sl Gaia} DR2 astrometric and photometric catalogues and the 2018 release of VHS photometric catalogue.  Fig. 21 of \citetalias{prime4} shows that L subdwarfs are distinctive from field dwarfs on the $M_G$ versus $J - K$ HRD.
 Fig. \ref{hrd} shows the modified $M_G$ vs $G - G_{RP}$ and $M_G$ vs $J - K$ HRD of \citetalias{prime4}. 20 known L subdwarfs observed in {\sl Gaia} DR2 can be separated from M and L dwarfs by the dashed line in Fig. \ref{hrd}, defined by equations \ref{mjk1} and \ref{mjk2}. I also applied two cuts defined by equations \ref{gjk1} and \ref{gjk2} in the $G-J$ vs $J-K$ space to removed main sequence stars due to mismatches between {\sl Gaia} and VHS.  

\begin{eqnarray}
\label{mjk1}
M_{G} > 10.0 (J - K) + 7.0 \\
\label{mjk2}
M_{G} > 1.25 (J - K) + 13.5625 \\
\label{gjk1}
J - K < G - J - 3.0 \\
\label{gjk2}
J - K < 0.2 (G-J) 
\end{eqnarray}

\begin{eqnarray}
\label{mjk1v}
M_{G} > 10.0 (J - Ks) + 7.5 \\
\label{mjk2v}
M_{G} > 1.25 (J - Ks) + 13.625 \\
\label{gjk1v}
J - Ks < G-J - 2.95 \\
\label{gjk2v}
J - Ks < 0.2 (G-J) - 0.05 
\end{eqnarray}

Note that the NIR photometry in equations \ref{mjk1} -- \ref{gjk2} is in MKO system, e.g. the UKIRT Infrared Deep Sky Survey \citep[UKIDSS;][]{lawr07}. UKIDSS and VISTA have very similar $J$ band photometry but slightly different in $K/Ks$ bands. Fig. 7 of \citetalias{prime6} shows that MKO $K$ photometry of L0--6 dwarfs can be transformed to VISTA $Ks$ photometry by $K - Ks = -0.05$. The criteria to selection L subdwarfs with {\sl Gaia} and VISTA are described in equations \ref{mjk1v} -- \ref{gjk2v}. 

\begin{eqnarray}
\label{grp}
G - G_{RP} > 1.3 \\
G - J < 6.0 \\
13.0 < J < 18.5 \\
J_{Err}< 0.1 \\
Ks > 12.0 \\ 
\label{kse}
Ks_{Err} < 0.2 
\end{eqnarray}

\begin{eqnarray}
\label{b5d}
|b| > 15\degr \\
\label{plx}
parallax > 5.0 \\
\label{plxerr}
parallax/parallax\_error > 5.0 \\
\label{vtan}
V_{tan} > 100~ km/s 
\end{eqnarray}

Equations \ref{grp} -- \ref{kse} show six extra photometric criteria applied in my selection. The region within 15\degr~from the Galactic plane is avoided (equation \ref{b5d}). These 20 known L subdwarfs in {\sl Gaia} DR2 are within 120 pc, thus I only selected objects within 200 pc in {\sl Gaia} DR2 (equations \ref{plx} and \ref{plxerr}).  
Fig. 23 of \citetalias{prime4} shows that L subdwarfs and dwarfs are relatively well separated by their tangential velocity ($V_{tan}$) at 100 km s$^{-1}$.  
Therefore, I selected objects with $V_{tan} > 100$ km s$^{-1}$ (equation \ref{vtan}).

I carried out a by-eye image check of objects survived my selection criteria (equation \ref{mjk1}--\ref{vtan}). 88 objects were left on the list of L subdwarf candidates. 
Three of them are previous known UCSDs: WISEA J001450.17$-$083823.4 \citep{kirk14,luhm14}; ULAS J033351.10+001405.8 \citep{lodi12}; and SSSPM J10130734$-$1356204 \citep{scho04}, and were classified as L subdwarfs in \citetalias{prime1}. 
Note that most of these 20 known L subdwarfs in {\sl Gaia} DR2 are in the northern sky, and the VHS is a southern sky survey. Some L subdwarfs could be missed in the cross-matching between {\sl Gaia} and VHS due to high proper motion and long baseline.

\subsection{Binaries of L subdwarf candidates}
I cross-matched these 88 L subdwarf candidates with a sample of about 1.8 million {\sl Gaia} DR2 sources within 200 pc and off the Galactic plan ($|b| > 15\degr$) by proper motion within a separation of 3 arcmin. I allowed an match error of 5 mas yr$^{-1}$ for both $\mu_{\rm RA}$ and $\mu_{\rm Dec}$, and found two wide binaries with common proper motions.  Components of each pair also have very close parallax distances (see Table \ref{prop}). I also cross-matched these 20 L subdwarfs observed in {\sl Gaia} DR2 \citepalias[table 9;][]{prime4} with these 1.8 million {\sl Gaia} DR2 sources in the same way, but did not find any binary. 

Gaia DR2 4818823636756117504 (Gaia J045238.82$-$361001.3; hereafter Gaia J0452$-$36A) and Gaia DR2 4818823808553134592 (Gaia J045245.87$-$360843.8; hereafter Gaia J0452$-$36B) are at a distance of 137.27$^{+0.68}_{-0.67}$ pc separated by 115.3 arcsec corresponding to a projected separation of 15828$\pm$78 au. 
Gaia DR2 3466916670990633088 (Ruiz 440-469; hereafter Ruiz 440-469A) and Gaia DR2 3466916778361936000 (Gaia J115626.32$-$322227.1; hereafter Ruiz 440-469B) are at a distance of 112.94$^{+3.77}_{-3.53}$ pc separated by 77.5 arcsec, corresponding to a projected separation of 8794$^{+292}_{-174}$ au. Fig. \ref{vhsj} shows the VISTA $J$-band images of Gaia J0452$-$36AB and Ruiz 440-469. 

The statistic probability that these two pairs of stars separated by a few arcmins have such very close proper motion and distance by random chance is very tiny and negligible \citep[e.g,][]{zha10}. The probability is much smaller if both stars of such a pair are metal-poor or kinematically associated with the halo or thick disc.

\begin{figure}
\begin{center}
   \includegraphics[width=\columnwidth]{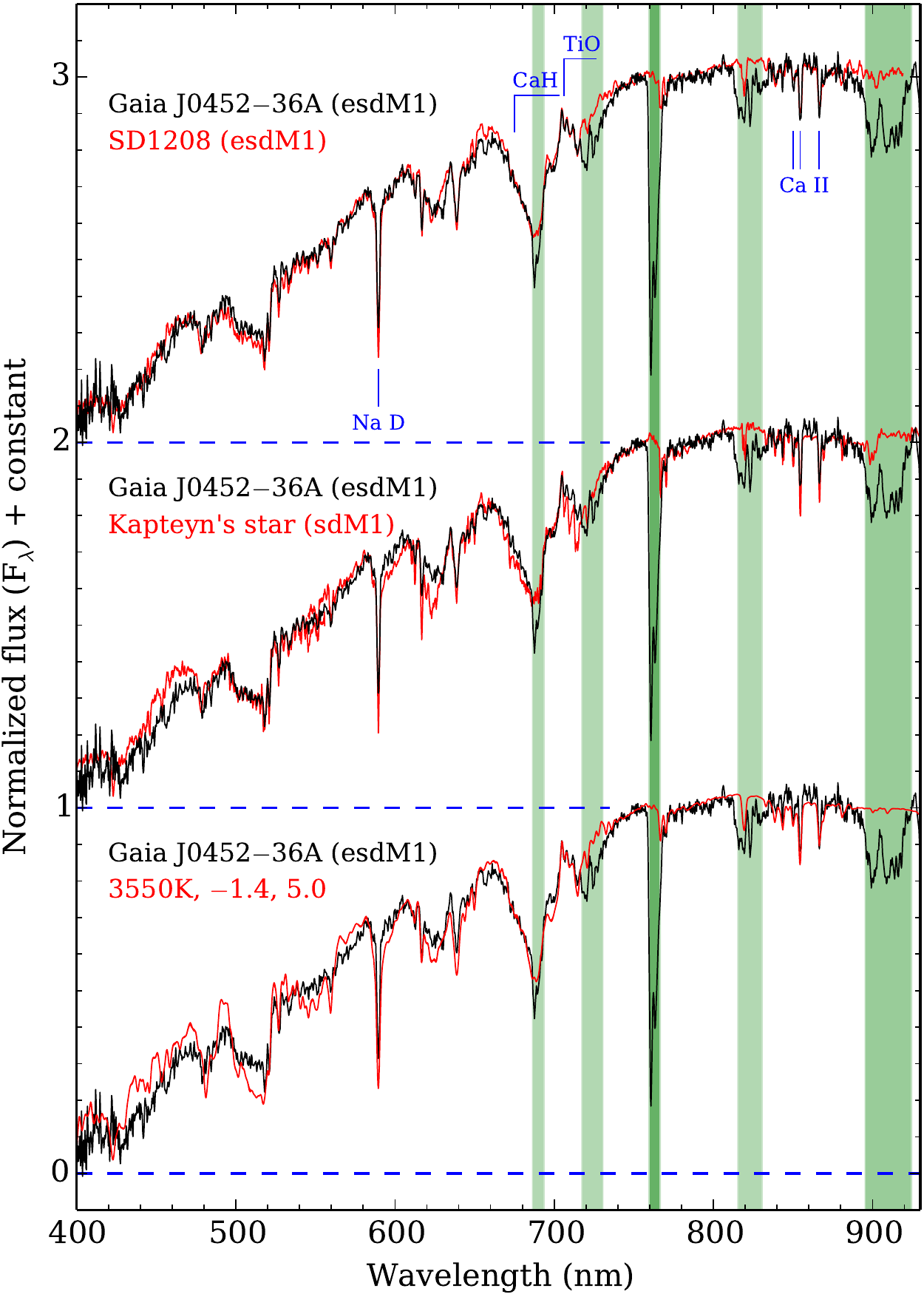}
\caption[]{ACAM spectrum of Gaia J0452$-$36A (without telluric correction) compared to SDSS spectrum of SD1208 (esdM1, telluric corrected), X-shooter spectrum of Kapteyn's star (sdM1, telluric corrected), and a BT-Settl model spectrum with $T_{\rm eff}$ = 3550 K, [Fe/H] = $-$1.4, and log $g$ = 5.0. The SDSS spectrum of SD1208 is smoothed by 5 pixels. The X-shooter spectrum of the Kapteyn's star is smoothed by 51 pixels. Telluric absorptions are indicated with shaded green bands. Lighter and thicker shaded bands indicate regions with weaker and stronger telluric effects.}
\label{j0252a}
\end{center}
\end{figure}

\begin{table*}
 \centering
  \caption[]{Properties of Gaia J0452$-$36AB and Ruiz 440-469AB wide binary systems.}
\label{prop}
  \begin{tabular}{l c c c c}
\hline
Component & Gaia J0452$-$36A &  Gaia J0452$-$36B  & Ruiz 440-469A &  Ruiz 440-469B  \\	
\hline 
Gaia DR2 & 4818823636756117504 & 4818823808553134592 & 3466916670990633088 & 3466916778361936000 \\
 $\alpha$ (2015.5)  & $04^{\rm h}52^{\rm m}38\fs82$ & $04^{\rm h}52^{\rm m}45\fs87$ & $11^{\rm h}56^{\rm m}22\fs77$  & $11^{\rm h}56^{\rm m}26\fs32$ \\
 $\delta$ (2015.5)  &  $-36\degr10\arcmin01\farcs3$ &  $-36\degr08\arcmin43\farcs8$  &  $-32\degr23\arcmin30\farcs2$ & $-32\degr22\arcmin27\farcs1$ \\
$G$ &  15.933 & 20.120 & 19.135 & 20.398 \\
$G_{\rm BP}$ &  16.892 & 21.080 & 19.407 & 21.612 \\
$G_{\rm RP}$ &  14.962 & 18.615 & 18.654 & 18.737 \\
$J$ (VHS) &  13.687$\pm$0.002 & 16.437$\pm$0.008 & 18.150$\pm$0.054 & 16.071$\pm$0.009 \\
$Ks$ (VHS) &  12.985$\pm$0.003 & 15.954$\pm$0.038 & 17.866$\pm$0.171 & 15.311$\pm$0.017 \\
$W1$ ({\sl WISE}) &  12.865$\pm$0.024 & 15.571$\pm$0.037 & 17.334$\pm$0.133 & 15.072$\pm$0.034 \\
$W2$ ({\sl WISE}) &  12.701$\pm$0.024 & 15.234$\pm$0.068 & 17.123$\pm$0.404 & 14.750$\pm$0.060 \\
$\pi$ (mas) & 7.285$\pm$0.036 & 7.134$\pm$0.506 & 8.854$\pm$0.286 & 8.851$\pm$0.818 \\
Distance (pc) &  137.27$^{+0.68}_{-0.67}$ & 140.17$^{+10.71}_{-9.29}$ & 112.94$^{+3.77}_{-3.53}$ & 112.98$^{+11.50}_{-9.55}$   \\
$\mu_{\rm RA}$ (mas yr$^{-1}$) &   148.45$\pm$0.06 & 147.52$\pm$0.79 & $-$206.72$\pm$0.61 & $-$209.43$\pm$1.93  \\
$\mu_{\rm Dec}$ (mas yr$^{-1}$) &  $-$168.70$\pm$0.07 &$-$168.07$\pm$1.00 & $-$41.85$\pm$0.29 & $-$44.21$\pm$0.99 \\
$V_{tan}$ (km s$^{-1}$) &  146.21$^{+0.72}_{-0.71}$ &  148.58$^{+11.35}_{-9.85}$ & 112.91$^{+3.77}_{-3.53}$ &  114.63$^{+11.67}_{-9.69}$ \\
RV (km s$^{-1}$) &  34$\pm$23 & --- & --- & ---  \\ 
$U$ (km s$^{-1}$) & $65\pm$9 & --- & --- & ---  \\ 
$V$ (km s$^{-1}$) & $-129\pm$15 & --- & --- & ---  \\ 
$W$ (km s$^{-1}$) & $42\pm$14 & --- & --- & ---  \\ 
Spectral type & esdM1 & esdL0 & DA WD & M8 \\
$T_{\rm eff}$ (K) & 3550$\pm$100 & 2600$\pm$100 & --- & ---  \\
${\rm [Fe/H]}$ & $-$1.4$\pm$0.2 & $-$1.4$\pm$0.2 & --- & --- \\
log $g$ & 5.0$\pm$0.2 & 5.5$\pm$0.2 & --- & --- \\
Mass (M$_{\sun}$) & 0.151$^{+0.029}_{-0.019}$ & 0.0855$^{+0.0014}_{-0.0010}$ & --- & ---\\
\hline 
Separation (arcsec) &  \multicolumn{2}{c}{115.3}   &  \multicolumn{2}{c}{77.5}  \\
Projected separation (au) &  \multicolumn{2}{c}{15828$\pm$78}  &  \multicolumn{2}{c}{8794$^{+292}_{-174}$}  \\
Projected separation ($r_{\rm J}$) &  \multicolumn{2}{c}{0.092}   &  \multicolumn{2}{c}{---}  \\
$-U$ (J) &  \multicolumn{2}{c}{$1.44\times10^{33}$} & \multicolumn{2}{c}{---}  \\
\hline
\end{tabular}
\end{table*}

\begin{table*}
 \centering
  \caption[]{Summary of the characteristics of the spectroscopic observations made with WHT and VLT.   }
\label{tobs}
  \begin{tabular}{l r c c c c c c r r}
\hline
    Name  & SpT & UT date & Telescope & Instrument/grism & Slit & Seeing  & Airm  & $T_{\rm int}$ (VIS) & $T_{\rm int}$ (NIR)      \\
     & & & & &  (arcsec) &  (arcsec) & & (s)~~~~ & (s)~~~~\\   
\hline
Gaia J0452$-$36A & esdM1 & 2019-02-03 & WHT & ACAM/V400 & 1.0 & 1.3 & 2.35 &  600 & ---   \\
Gaia J0452$-$36B & esdL0  & 2019-02-03 & WHT & ACAM/V400 & 1.0 & 1.3 & 2.34 &  1200 &---  \\
Ruiz 440-469B & M8  & 2019-02-03 & WHT & ACAM/V400 & 1.0 & 1.3 & 2.08 & 1200 & --- \\
SSSPM 1013$-$13 & usdL0 & 2019-02-02 & WHT & ACAM/V400 & 1.0 & 1.3 & 1.41 &  300 &--- \\
Kapteyn's star & sdM1 & 2016-09-14 & VLT & X-shooter & 1.2 & 1.18 & 1.31 & $2 \times 5$ & $2 \times 5$ \\
WI0459  & esdM6 & 2016-01-23 & VLT & X-shooter & 1.2 & 1.23 & 1.33  & $4 \times 290$ & $4 \times 300$ \\
\hline
\end{tabular}
\end{table*}

\section{Spectroscopy}
\label{sobs}
Optical spectra of Gaia J0452$-$36AB and Ruiz 440-469B were observed as backup targets with the Auxiliary-port CAMera \citep[ACAM;][]{benn08} on the William Herschel Telescope (WHT) under the programme 95-WHT10/19A (PI: M. C. G\'{a}lvez Ortiz). The Kapteyn's star and WISEA J045921.22+154059.2 \citep[WI0459;][]{kirk16} were included in following discussions in Section \ref{smc}. Therefore, I present their optical to NIR spectra observed with the X-shooter \citep{vern11} on the Very Large Telescope (VLT). 

\subsection{William Herschel Telescope}
\begin{figure}
\begin{center}
   \includegraphics[width=\columnwidth]{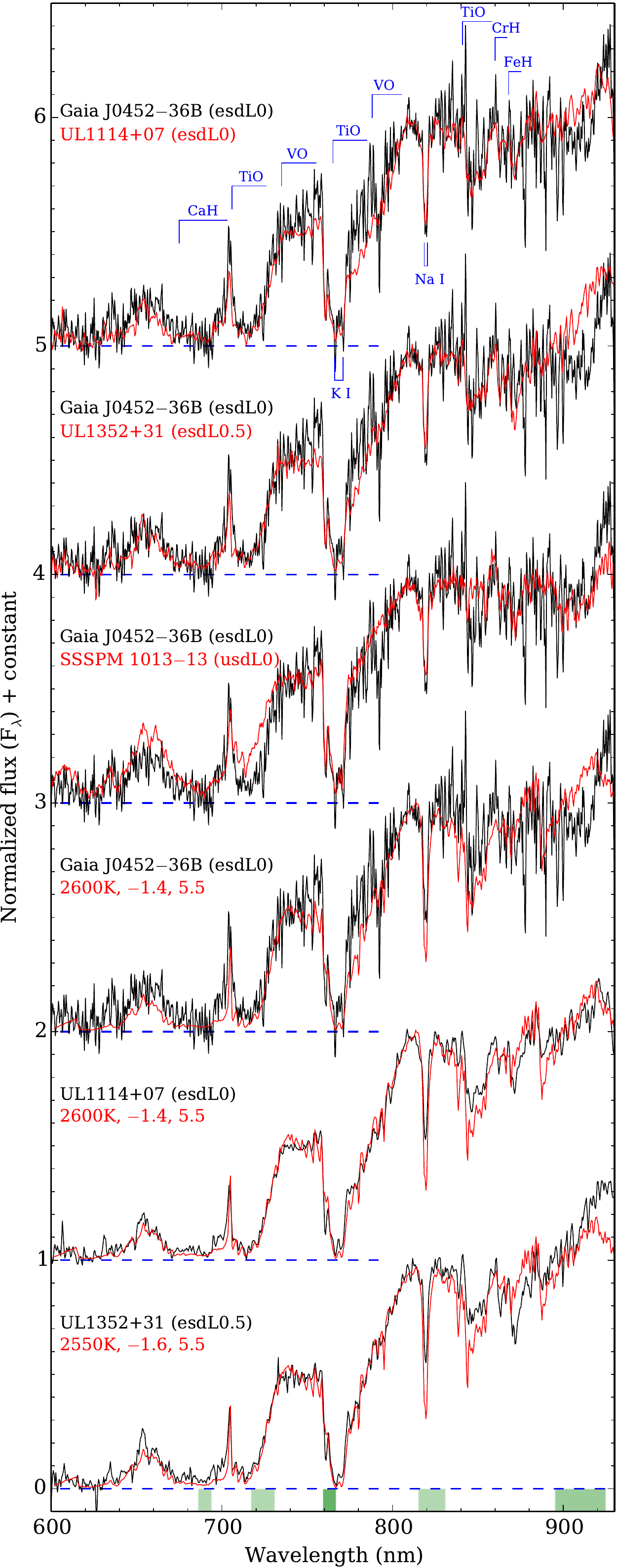}
\caption[]{The ACAM spectrum of Gaia J0452$-$36B compared to UL1114+07 \citetalias[esdL0;][]{prime4}, UL1352+31 \citetalias[esdL0.5;][]{prime4}, SSSPM 1013$-$13 (usdL0), and a BT-Settl model spectrum. Telluric absorptions are not corrected (shaded green bands).  
}
\label{j0252b}
\end{center}
\end{figure}

\begin{figure}
\begin{center}
   \includegraphics[width=\columnwidth]{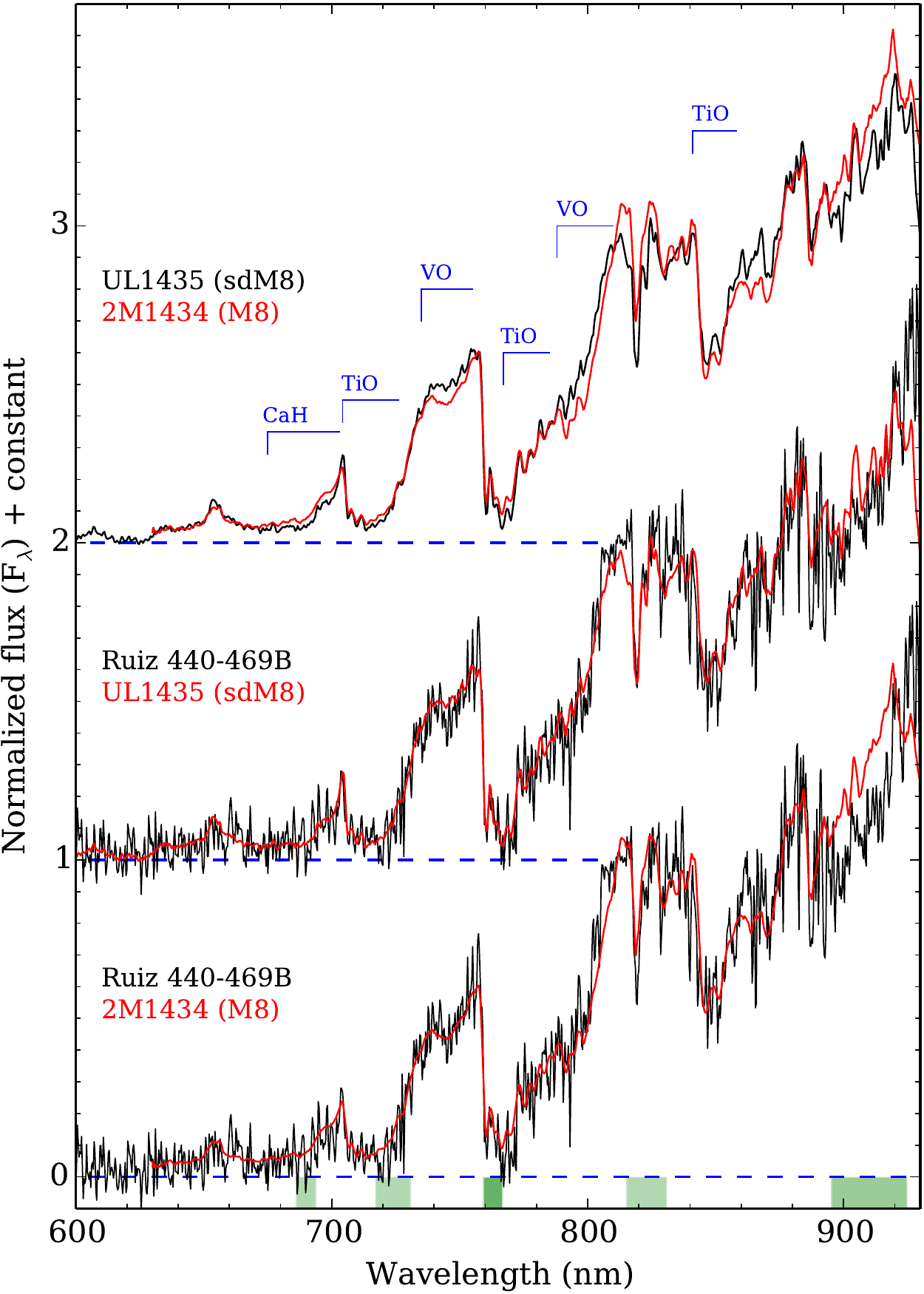}
\caption[]{The optical spectrum of Ruiz 440-469B compared to those of UL1435 \citepalias[sdM8;][]{prime4} and 2M1434 \citep[M8;][]{kirk99}. Telluric absorptions are not corrected (shaded green bands).}
\label{j1156b}
\end{center}
\end{figure}

Gaia J0452$-$36AB were observed with ACAM on 2019 February 3, under the seeing of 1.3 arcsec, and air mass of 2.35 for Gaia J0452$-$36A and 2.34 for Gaia J0452$-$36B (Figs \ref{j0252a} and \ref{j0252b}). The ACAM wavelength coverage is 390--930 nm. A V400 grism and 1 arcsec slit were used for the ACAM observation, providing a resolving power of 570 at 750 nm. The integration times for Gaia J0452$-$36AB are 600 and 1200 s, respectively. The peak signal-to-noise ratio (S/N) of their spectra is 74 at 750 nm for Gaia J0452$-$36A and 21 at 810 nm for Gaia J0452$-$36B. 
Ruiz 440-469B was observed on 2019 February 3, under the seeing of 1.3 arcsec, air mass of 2.08, with integration times of 1200 s (Fig. \ref{j1156b}). The peak S/N of its spectrum is about 21 at 810 nm. I also observed a known L0 ultra subdwarf (usdL0), SSSPM J10130734$-$1356204 (SSSPM 1013$-$13; \citealt{scho04}; \citetalias{prime1}). 
ACAM observational characteristics are summarised in Table \ref{tobs}.

These ACAM spectra were reduced with a standard {\scriptsize IRAF} package\footnote{IRAF is distributed by the National Optical Observatory, which is operated by the Association of Universities for Research in Astronomy, Inc., under contract with the National Science Foundation.}. 
The flux calibration was achieved with an F8V standard star HD84937 observed in a 1.0 arcsec slit, seeing of 1.0, and air mass of 1.04 on 2019 February 20. Telluric absorptions in ACAM spectra are not corrected.  

\subsection{Very Large Telescope}
\begin{figure}
\begin{center}
   \includegraphics[width=\columnwidth]{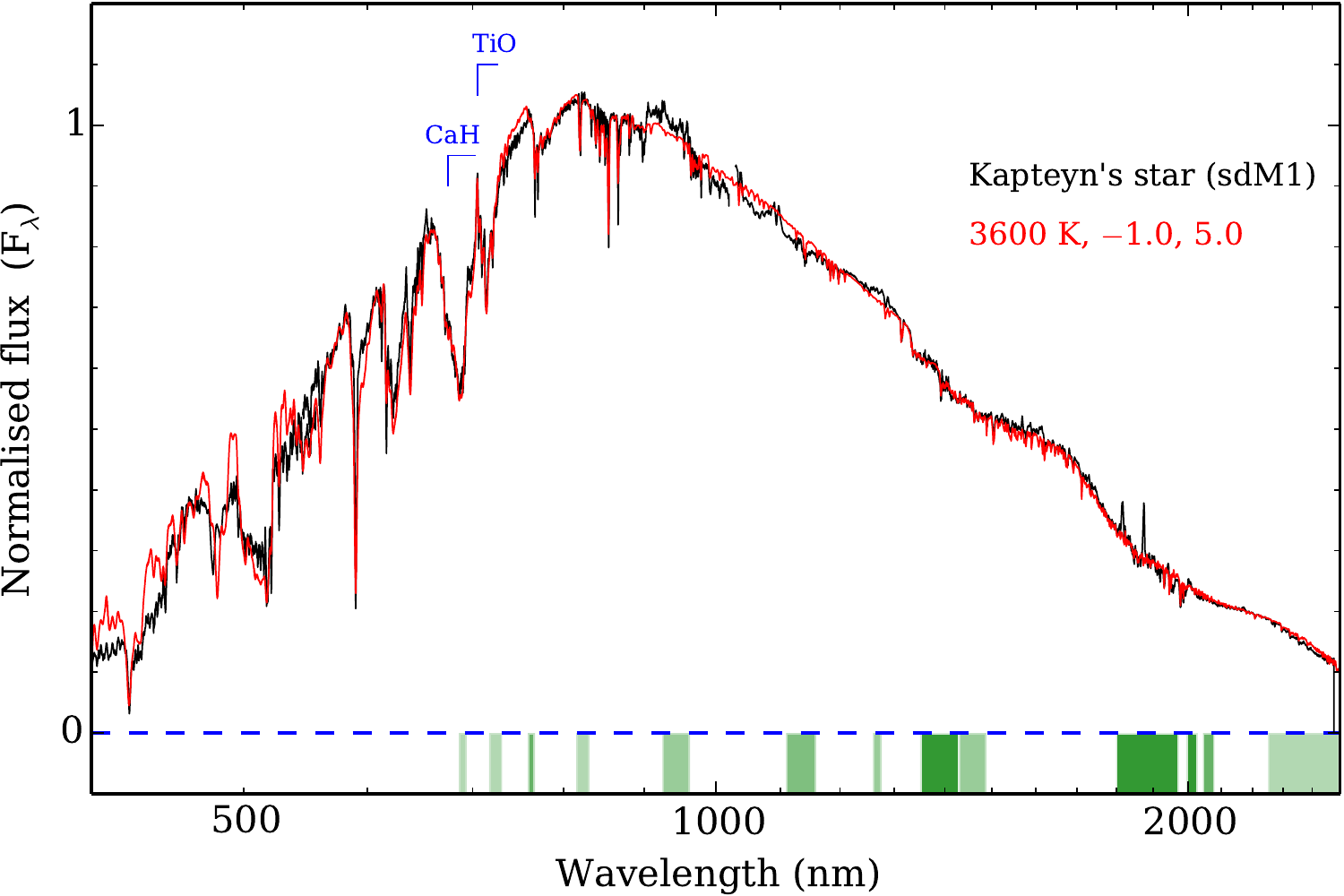}
\caption[]{The X-shooter spectrum (smoothed by 51 pixels) of Kapteyn's star and its best-fitting BT-Settl model spectrum with $T_{\rm eff}$ = 3600 K, [Fe/H] = $-$1.0, and log $g$ = 5.0.}
\label{gj191}
\end{center}
\end{figure}

The X-shooter spectrum of the Kapteyn's star (sdM1) was observed in a 1.2 arcsec slit under the seeing of 1.18 arcsec and air mass of 1.31 on 2016 September 14. The total exposure was brake up into two single integrations of 5 s with a AB nod separated by 5 arcsec on the detector. The spectrum was reduced to a flux-calibrated 2D spectrum with the ESO (European Southern Observatory) Reflex \citep{freu13}, then extracted to a 1D spectrum with the {\scriptsize IRAF APSUM}. The spectrum displayed in Fig. \ref{gj191} is smoothed by 51 pixels, which increased the S/N by $\sim$ 7 times. Telluric correction was achieved using a A0 V telluric standard (HD 216009) which was observed just before the target at the air mass of 1.41. 

\begin{figure}
\begin{center}
   \includegraphics[width=\columnwidth]{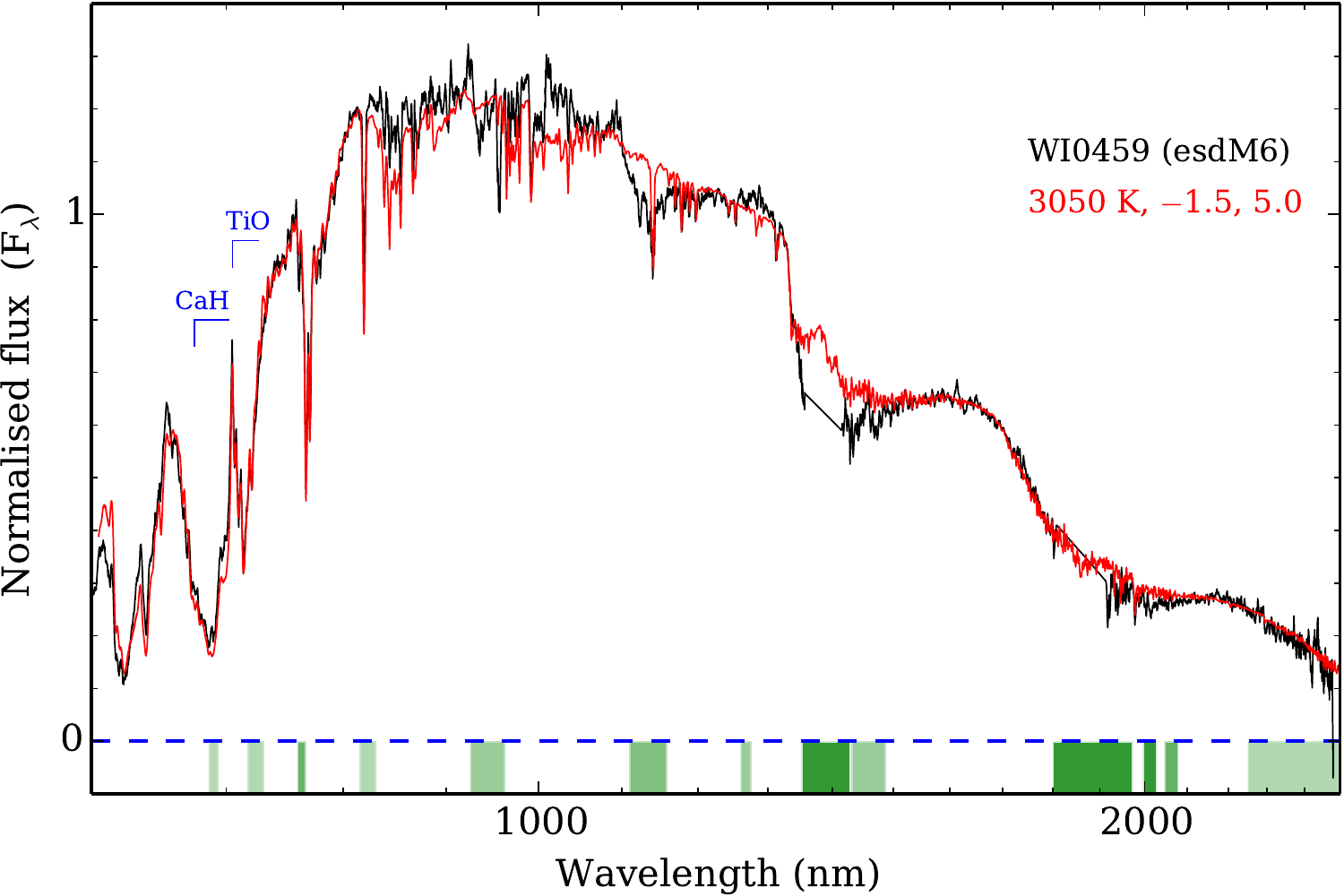}
\caption[]{The optical-to-NIR spectrum of WI0459 observed with the X-shooter (smoothed by 101 pixels in the VIS and 51 pixels in the NIR). Its best-fitting BT-Settl model has $T_{\rm eff}$ = 3050 K, [Fe/H] = $-$1.5, and log $g$ = 5.0. }
\label{wi0459}
\end{center}
\end{figure}

The X-shooter spectrum of WI0459 was observed in a 1.2 arcsec slit under the seeing of 1.23 arcsec and air mass of 1.33 on 2016 January 23. The total exposure was brake up into four single integrations of 290 s for the visible (VIS) arm and 300 s for the NIR arm in an ABBA nodding mode. The reduction procedure is the same as for the Kapteyn's star. The spectrum of WI0459 has an S/N of 63 at 825nm, and 40 at 122 and 135 nm. The spectrum displayed in Fig. \ref{wi0459} is smoothed by 101 pixels in the VIS and 51 pixels in the NIR, which increased the S/N by 10 and 7 times, respectively. Telluric correction was achieved using a A1 II telluric standard (HD 40335) which was observed right after the target at the air mass of 1.18.

\begin{figure}
\begin{center}
   \includegraphics[width=\columnwidth]{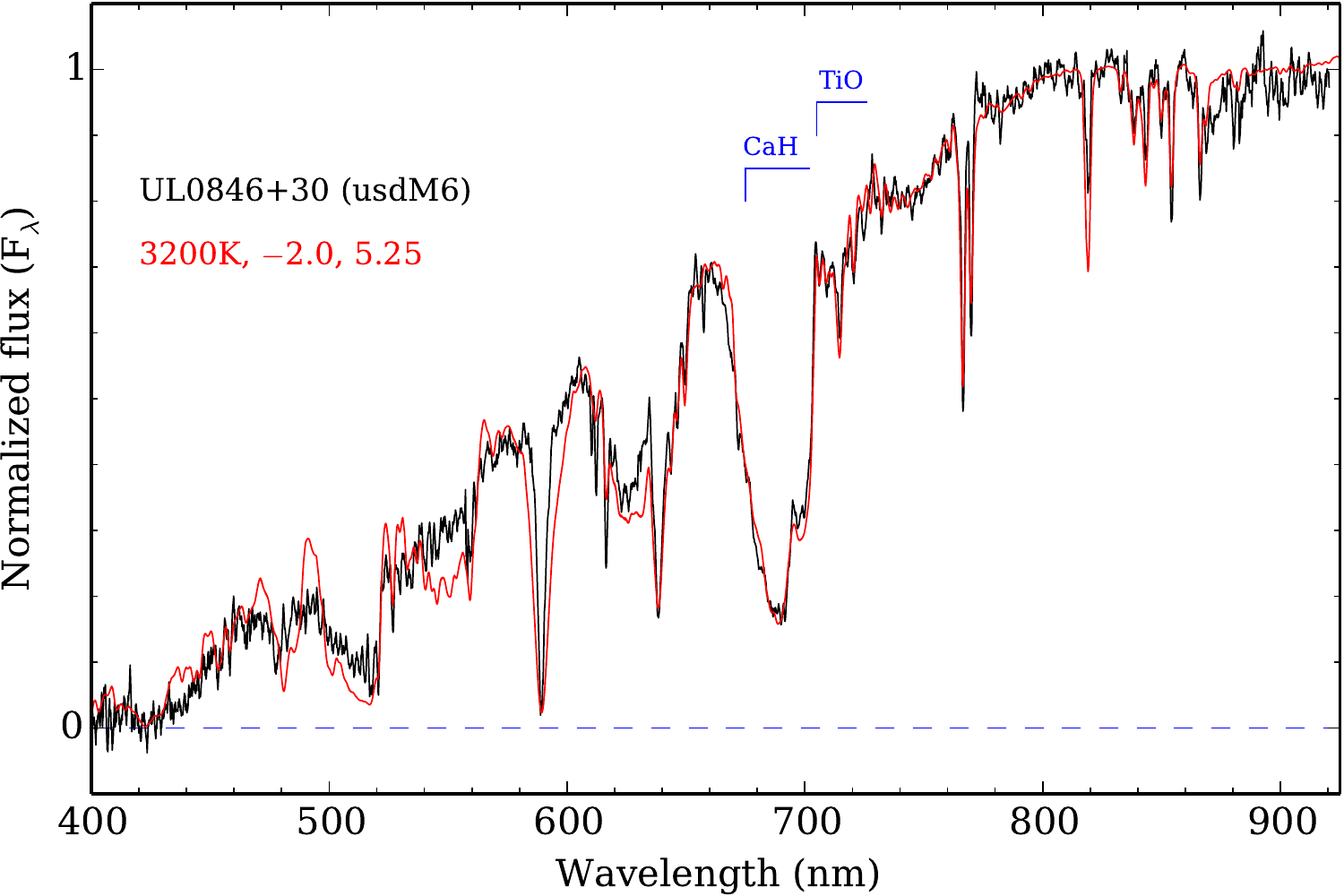}
\caption[]{The SDSS optical spectrum of SD0846+30 and its best-fitting BT-Settl model with $T_{\rm eff}$ = 3200 K, [Fe/H] = $-$2.0, and log $g$ = 5.25. The SDSS has an average resolving power of about 2000 and is smoothed by 3 pixels. }
\label{sd0846}
\end{center}
\end{figure}

\section{Characterization}
\label{scha}

\subsection{Gaia J0452--36AB}
Gaia J0452$-$36A is located off the dwarf main sequence on the $G - G_{\rm RP}$ versus $M_G$ HRD (Fig. \ref{hrd}). It has slightly fainter $M_G$ and bluer $G - G_{\rm RP}$ colour than the Kapteyn's star and located between two components of the esdK5+esdM5.5 wide binary G224-58AB \citep{zha13,pavl15}. Gaia J0452$-$36A is below a gap among M dwarfs \citep{jao18} associated with mixing of $^3$He that reduced the nuclear fusion rate in a narrow mass range, around 0.34--0.36 M$_{\sun}$ for solar metallicity \citep{bara18,macd18}. Gaia J0452$-$36B is not separated from the main sequence in the $M_G$ versus $G - G_{\rm RP}$ HRD. Because UCSDs have both bluer $G - G_{\rm RP}$ and brighter $M_G$ than UCDs with equivalent mass, thus located to the upper left of UCDs and associated with higher mass red dwarfs or UCDs on the main sequence \citepalias{prime4}. Fig. \ref{hrd} (right-hand panel) shows that Gaia J0452$-$36B has distinctive $J-K$ colour from dwarfs. Early-type M subdwarfs have similar $J-K$ colour as dwarfs \citepalias[e.g. fig. 1,][]{prime1}, thus Gaia J0452$-$36A appears like associated with dwarfs on the the main sequence in the $M_G$ versus $J-K$ HRD.

\subsubsection{Spectral classification}
Fig. \ref{j0252a} shows that the optical spectrum of Gaia J0452$-$36A fits well to an esdM1 subdwarf, SDSS J120854.81+284031.0 (SD1208, telluric corrected), except the telluric regions. The $\zeta_{\rm CaH/TiO}$ index of Gaia J0452$-$36A is 0.306 which is in the range of esdM subclass in the classification scheme of \citet{lepi07}. Therefore, I classified Gaia J0452$-$36A as an esdM1 subdwarf. Note that the wavelength ranges used to define the $\zeta_{\rm CaH/TiO}$ index are outside of telluric regions \citep[table 1 of]{lepi07}. Gaia J0452$-$36A also shown a weaker TiO absorption around 705--718 nm than the Kapteyn's star which is an sdM1 subdwarf. The Kapteyn's star has a metallicity of [Fe/H] $\approx -0.99\pm0.04$ \citep{wool05} and is relatively metal-poor in the sdM subclass \citepalias[e.g. table 8;][]{prime1}. 
Note that the TiO absorption at 718--726 nm in Gaia J0452$-$36A is contaminated by telluric absorption.  

Gaia J0452$-$36A has a radial velocity (RV) of 34$\pm$23 km s$^{-1}$. To measure the RV of Gaia J0452$-$36A, I used SD1208, which has an RV of $-91.08\pm3.23$ km s$^{-1}$, as a reference. Strong common absorption lines (e.g. Na D, Ca II) in Gaia J0452$-$36A and SD1208 were used to measure their RV difference. Then the barycentric velocity was corrected for Gaia J0452$-$36A. I also calculated the space velocity of Gaia J0452$-$36A based on its astrometry from {\sl Gaia} DR2 and RV measured from its ACAM spectrum. The space velocity of Gaia J0452$-$36A ($U = 65\pm$9 km s$^{-1}$; $V = -129\pm$15 km s$^{-1}$; and $W = 42\pm$14 km s$^{-1}$) is typical for halo population. The halo membership of Gaia J0452$-$36AB is robust considering its metallicity ([Fe/H] = $-$1.4; see Section \ref{spro}) is also typical for halo population.

\citetalias{prime1} presented a classification scheme for L subdwarfs, and classified L subdwarfs into three subclasses: L subdwarf (sdL), L extreme subdwarf (esdL), and L ultra subdwarf (usdL). Metallicity of each L subclass consistent across subtypes and consistent with that of M0--3 subdwarf subclasses defined by \citet{lepi07}.  
Fig. \ref{j0252b} shows the optical spectrum of Gaia J0452$-$36B compared to known L subdwarfs. Spectral types of these known L subdwarfs are based on the classification scheme of \citetalias{prime1}. The spectrum of Gaia J0452$-$36B fits well to that of the esdL0 type ULAS J111429.54+072809.5 (UL1114+07) and esdL0.5 type ULAS J135216.31+312327.0 \citepalias[UL1352+31;][]{prime4},  particularly at CaH, TiO, VO, CrH, and FeH absorption bands at 670--740 and 790--880 nm, which are sensitive to temperature and metallicity. The flux of Gaia J0452$-$36B at around 750 and 770--790 nm looks higher than UL1114+07 and UL1352+31. This is not a real feature but caused by contamination of noise. Gaia J0452$-$36B clearly has slightly stronger TiO absorption at around 720 and 850 nm (sensitive to metallicity) than the usdL0 type SSSPM 1013$-$13. As a result, I classified Gaia J0452$-$36B as an esdL0 subdwarf. Note that the TiO absorption around 720 nm in Gaia J0452$-$36B is partially in a telluric region. However, the telluric absorption is much weaker than the TiO absorption around 720 nm, and has relatively small impact on the TiO absorption band around 720 nm. 

\subsubsection{Physical properties}
\label{spro}
I fitted the optical spectra of Gaia J0452$-$36AB with BT-Settl model spectra \citep{alla14}. Figs \ref{j0252a} and \ref{j0252b} show the best-fitting model spectra of Gaia J0452$-$36AB, which both have [Fe/H] = $-$1.4. Gaia J0452$-$36A has $T_{\rm eff}$ = 3550 K and log $g$ = 5.0. Gaia J0452$-$36B has $T_{\rm eff}$ = 2600 K and log $g$ = 5.5. The fitting procedure is described in previous papers of this series. I estimated masses of Gaia J0452$-$36AB based on the 10 Gyr iso-mass contours predicted by evolutionary models \citep{bara97,chab97} plotted in the $T_{\rm eff}$ versus [Fe/H] space \citepalias{prime2,prime3}. Gaia J0452$-$36AB have masses of 0.151$^{+0.029}_{-0.019}$ and 0.0855$^{+0.0014}_{-0.0010}$ M$_{\sun}$, respectively. 

The Kapteyn's star has $T_{\rm eff} = 3570\pm160$ K and [Fe/H] = 0.99$\pm$0.04 according to \citet{wool05}. Fig. \ref{gj191} shows that the X-shooter spectrum of the Kapteyn's star fitted well to a BT-Settl model spectrum with [Fe/H] = $-$1.0, $T_{\rm eff}$ = 3600 K, and log $g$ = 5.0. Gaia J0452$-$36A has a weaker TiO absorption band around 710 nm than Kapteyn's star (Fig. \ref{j0252a}), thus certainly have lower metallicity than Kapteyn's star. With $T_{\rm eff}$ = 3600 K and [Fe/H] = $-$1.0, Kapteyn's star would have a mass of 0.21$^{+0.06}_{-0.04}$ M$_{\sun}$ \citepalias[fig. 9,][]{prime2}, which is significant lower than estimates in the literature (0.274 M$_{\sun}$, \citealt{koto05}; 0.281 M$_{\sun}$ \citealt{angl14}).

Two known M subdwarfs, WI0459 \citep{kirk16} and SDSS J084648.88+302801.7 \citep[SD0846+30;][]{zha13}, are selected for the discussion in Section \ref{smc}. I observed an optical to NIR spectrum of WI0459 with the X-shooter (Fig. \ref{wi0459}). The optical spectrum of SD0846+30 (Fig. \ref{sd0846}) is from the Sloan Digital Sky Survey \citep[SDSS;][]{york00}. The best-fitting BT-Settl model spectra have [Fe/H] = $-$1.5, $T_{\rm eff}$ = 3050 K, and log $g$ = 5.0 for the esdM6 type WI0459, and [Fe/H] = $-$2.0, $T_{\rm eff}$ = 3200 K, and log $g$ = 5.25 for the usdM6 type SD0846+30. 

\subsubsection{Binary properties}
Gaia J0452$-$36AB are located at a distance of 137.27$^{+0.68}_{-0.67}$ pc with an angular separation of 115.3 arcsec, that corresponds to a projected separation of 15828$\pm$78 au. The projected separation of the system is about 0.092 Jacobi radius ($r_{\rm J}$; tidal radius). This suggests that Gaia J0452$-$36AB is a gravitationally bound system as its separation is far smaller than the Jacobi radius at an age of around 10 Gyr \citep{jian10}. The binding energy ($-U$) of the system is about $1.44\times10^{33} J$.

\subsection{Ruiz 440-469AB}
Ruiz 440-469A is a cool DA white dwarf \citep{ruiz96} and associated with the white dwarf sequence on both HRD in Fig. \ref{hrd}. Ruiz 440-469B has slightly bluer $J-K$ colour than dwarfs on the main sequence. 

Fig. \ref{j1156b} shows the optical spectrum of Ruiz 440-469B compared to that of an sdM8 subdwarf, ULAS J143517.18$-$014713.1 \citepalias[UL1435;][]{prime4}, and an M8 dwarf 2MASSW J1434264+194050 \citep[2M1434;][]{kirk99}. The S/N of Ruiz 440-469B is not good enough to distinguish the difference between an M8 and sdM8 type by the VO absorption band around 800 nm. Ruiz 440-469B has slightly bluer $J-K$ colour than normal M dwarfs on the main-sequence (Fig. \ref{hrd}). The VISTA $J-Ks$ colour of Ruiz 440-469B (0.76$\pm$0.02) is slightly bluer than that of UL1435 (0.79$\pm$0.01). Therefore, Ruiz 440-469B is likely an old mildly metal-poor M8 dwarf. 

As a DA WD + M8 dwarf binary, Ruiz 440-469AB has larger total mass and closer projected separation (8794$^{+292}_{-174}$ au) than Gaia J0452$-$36AB.  Assuming that Ruiz 440-469AB have masses of 0.5 and 0.1 M$_{\sun}$, respectively. The projected separation of Ruiz 440-469AB would be around 0.067 $r_{\rm J}$. Therefore, Ruiz 440-469AB is also a gravitationally bound system. Properties of Ruiz 440-469AB are listed in Table \ref{prop}.

\begin{figure*}
\begin{center}
   \includegraphics[width=\textwidth]{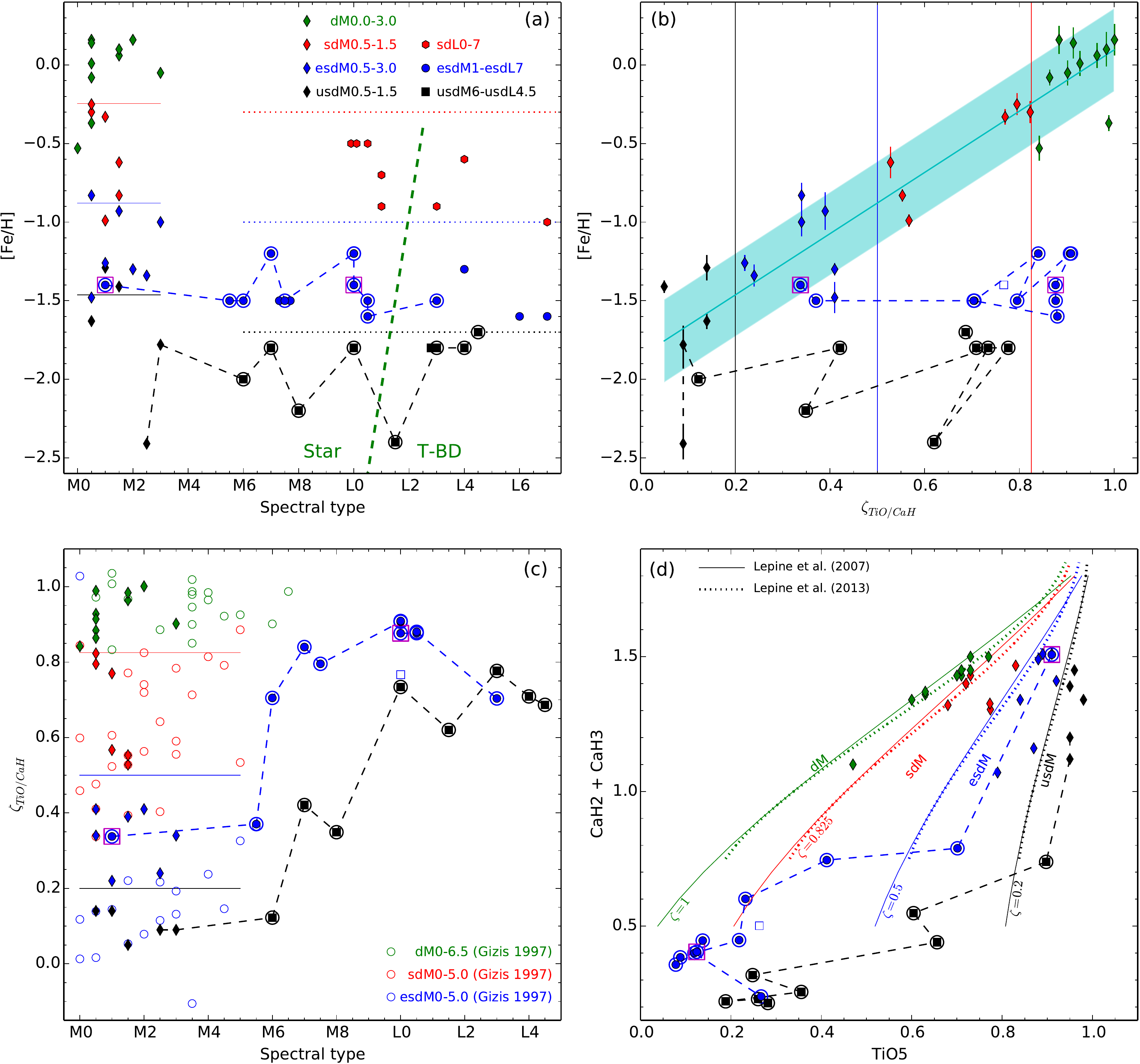}
\caption[]{Spectral type, metallicity, CaH2+CaH3, TiO5, and $\zeta_{\rm CaH/TiO}$ indices of Gaia J0452$-$36AB, and M and L subdwarfs with known metallicities. The spectral indices of Gaia J0452$-$36B (blue open square) are measured from its low S/N spectrum thus have large uncertainty. UL1114 have the same spectral type and metallicity as Gaia J0452$-$36B thus is used as a replacement. The distance between Gaia J0452$-$36B and UL1114 roughly shown the index uncertainty of Gaia J0452$-$36B. Gaia J0452$-$36A and UL1114 are highlighted with magenta squares. Diamonds are M0--3 dwarfs/subdwarfs with measured [Fe/H] \citep{wool05,wool06,wool09}. [Fe/H] of the rest M and L subdwarfs are from model spectral fittings (\citealt{pavl15}; \citetalias{prime1,prime2,prime3,prime5}; This paper). Objects with known [Fe/H] estimated from BT-Settl models and measured $\zeta_{\rm CaH/TiO}$ index (see Table \ref{tzeta}) are highlighted with open circles. Note that two pairs of them have the same spectral type and [Fe/H] thus are overlapped in panel (a). Objects in esd and usd subclasses with measured spectral indices are joined separately with dashed lines in the order of spectral types. This formed two metallicity trace lines along spectral types. The red, blue, and black dotted lines in panel (a) indicate the boundaries between sd, esd, and usd subclasses of UCSDs \citepalias{prime1}. A green dashed line in panel (a) indicates an empirical boundary between very low-mass stars and T-BDs \citepalias{prime2}. The red, blue, and black solid lines in panels (b--d) indicate boundaries between dM, sdM, esdM, and usdM defined by $\zeta_{\rm CaH/TiO}$ = 0.825, 0.5, and 0.2, respectively \citep{lepi07}. The dotted lines in panel (d) are for $\zeta_{\rm CaH/TiO}$ = 0.825, 0.5, and 0.2 defined in \citet{lepi13}. The green solid and dotted lines indicate the red dwarf sequence ($\zeta_{\rm CaH/TiO}$ = 1). The cyan line and shaded area in panel (b) are the polynomial fit and fitting rms of these M0--3 dwarfs/subdwarfs (Equation \ref{equ1}). The [Fe/H] boundaries between M0--3 subclasses of \citet{lepi07} are indicated with red, blue, and black solid lines in panels (a) which are derived from panel (b). The green, red, and blue open circles in panel (c) are M dwarfs, and sdM and esdM subdwarfs classified by \citet{gizi97}. }
\label{fsmz}
\end{center}
\end{figure*}

\begin{figure*}
\begin{center}
   \includegraphics[width=\textwidth]{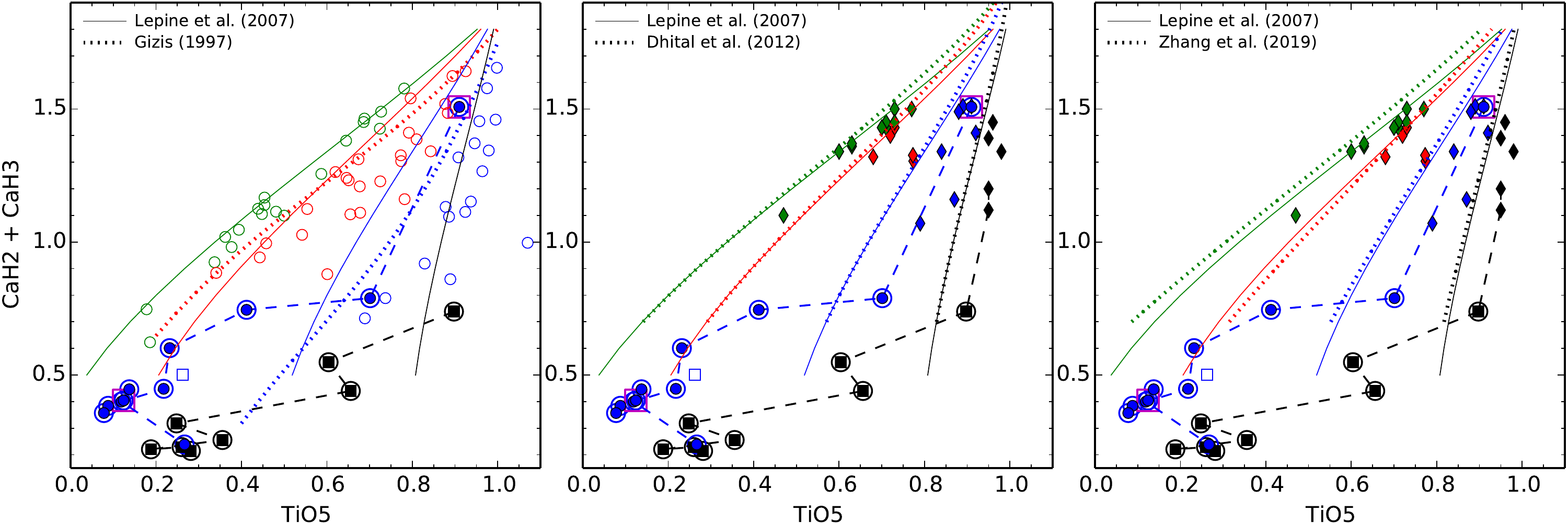}
\caption[]{TiO5 and CaH2+CaH3 indices of M subdwarf subclasses defined with the $\zeta_{\rm CaH/TiO}$ index by  \citet[left]{gizi97}, \citet[middle]{dhit12}, and \citet[right]{zhan19} comparied to those of \citet{lepi07}. The red and blue dotted lines on the left-hand panel are boundaries between dM, sdM, and esdM subclasses \citep{burg06} under the scheme of \citet{gizi97}. The other symbols are the same as in Fig. \ref{fsmz}. }
\label{ftiob}
\end{center}
\end{figure*}

\section{Metallicity consistency of M subclasses}
\label{smc}
\citetalias{prime1} (section 4.1) noticed that the metallicity is not consistent across mid-to-late M subtypes in each metallicity subclass (sdM, esdM, and usdM) that classified by the $\zeta_{\rm TiO/CaH}$ index \citep{lepi07}. Metallicity scale of late-type M subdwarfs are overestimated by the $\zeta_{\rm TiO/CaH}$ index, which is also not applicable for L subdwarfs. For example, the strength of the TiO absorption band at 850 nm is not a monotonous function of metallicity for late-type M and L subdwarfs (fig. 10 in \citetalias{prime1}; fig. 1 in \citetalias{prime5}).

UCSDs with well-constrained metallicities are required to test the metallicity consistency of M subdwarf classification. However, the metallicity of UCSDs are extremely difficult to measure directly by observation. First because they are faint and it is difficult to observe high-quality (S/N and resolution) spectra. Secondly, they emit most of their flux at longer wavelength, and have complex molecular absorptions. Only some M0--3 subdwarfs have direct metallicity measurements based on atomic lines at blue wavelength \citep[e.g.][]{wool05,wool06,wool09}. Thirdly, there are not enough wide F/G/K + M subdwarf binaries that can be used to calibrate the metallcity measurements of mid- and late-type M subdwarfs in the NIR \citep{roja12,newt15}. 

The metallicity consistency of M subclasses can be tested with wide binaries composed of early- and late-type M subdwarfs which have the same metallicity. However, wide binary systems contain UCSDs are extremely rare. First, because UCSDs belong to rare thick disc or halo populations. Secondly, UCSDs are faint and only these in the solar neighbourhood could be well observed. Thirdly, late-type M and L subdwarfs are in a smaller mass range on the mass function compared to late-type M and L dwarfs \citepalias[fig. 9;][]{prime2}. Lastly, some late-type M objects in the same metallicity range as early-type M subdwarfs are classified into late-type M dwarfs by the $\zeta_{\rm TiO/CaH}$ index \citepalias{prime1}. 

As a wide esdM1+esdL0 binary, Gaia J0452$-$36AB is the first ideal binary that can be used to conduct a test of the metallicity consistency in the classification of UCSDs. To provide a better view of correlation between metallicity and spectral indices that used in the classification of M subdwarfs across a broad spectral type range, I collected a sample of M and L subdwarfs with known metallicity (Table \ref{tzeta}). I studied the correlations between spectral type, metallicity ([Fe/H]), CaH and TiO indices, and $\zeta_{\rm TiO/CaH}$ index based on Gaia J0452$-$36AB and M and L subdwarfs in Table \ref{tzeta}. 

Fig. \ref{fsmz} (a) shows the correlations between spectral type and metallicity for M and L subdwarfs of different subclass. Metallicity subclasses are based on the classification schemes of \citet{lepi07} for early-type M subdwarfs, and \citetalias{prime1} for UCSDs. The metallicity boundaries between different subclasses of M0--3 subdwarfs are derived from the correlation between [Fe/H] and $\zeta_{\rm TiO/CaH}$ index of M0--3 subdwarfs in Fig. \ref{fsmz} (b). The metallicity boundaries of M0--3 sublcasses \citep{lepi07} and UCSD subclasses \citepalias{prime1} are roughly consistent. Objects with metallicity derived from BT-Settl models are highlighted with circles (see Table \ref{tzeta}). These esdM and esdL subdwarfs (including Gaia J0452$-$36AB) are joined with blue dashed lines in orders of spectral types (primary) and metallicity (secondary). Those usdM and usdL subdwarfs are also joined with black dashed lines. They formed two metallicity trace lines along spectral subtypes in the esdM/L and usdM/L subclasses.

\begin{table*}
 \centering
  \caption[]{The CaH2, CaH3, TiO5 spectral indices \citep{reid95}, $\zeta_{\rm TiO/CaH}$ index \citep{lepi07}, and metallicity of Kapteyn's star, and M and L subdwarfs that are used to form the esd and usd metallicity trace lines in Fig. \ref{fsmz}. Ref1 are references for the discoveries of these sources, and Ref2 are for spectral types (SpT) and [Fe/H]. 
  }
\label{tzeta}
  \begin{tabular}{l c l c c c c c c}
\hline
Name & Ref1 & SpT & [Fe/H] & Ref2 & CaH2 & CaH3 & TiO5 & $\zeta$	\\
\hline
Kapteyn's star & \citealt{kapt97} &  sdM1 & $-$0.99 & \citetalias{wool05} & 0.651 & 0.816 & 0.830 &	0.567 	\\
Gaia J045238.82$-$361001.3 & This paper &  esdM1 & $-$1.4 & This paper & 0.676 & 0.832 & 0.910 & 0.338	\\
Gaia J045245.87$-$360843.8 & This paper & esdL0  &  $-$1.4 & This paper & 0.101 & 0.400 & 0.263 & 0.767	\\
G 224-58B & \citealt{zha13} & esdM5.5  & $-$1.5 & \citealt{pavl15} & 0.287 & 0.502 & 0.701 & 0.370	\\
WISEA J045921.22+154059.2 & \citealt{kirk16} & esdM6 & $-$1.5 & This paper & 0.274 & 0.471 & 0.412 & 0.705 \\
LHS 377 & \citealt{gizi97} & esdM7  & $-$1.2 & \citetalias{prime1} & 0.205 & 0.396 & 0.232 & 0.563 \\
2MASS J01423153+0523285 & \citealt{burg04} & esdM7.5 & $-$1.5 & \citetalias{prime1} & 0.158 & 0.290 & 0.218 & 0.795	\\
WISEA J001450.17$-$083823.4 & \citealt{kirk14} & esdL0 & $-1.2$  &  \citetalias{prime1} & 0.126 & 0.258 & 0.088 & 0.907 \\
2MASS J16403197+1231068 & \citealt{burg04} & esdL0 & $-$1.2 & \citetalias{prime1} & 0.113 & 0.244 & 0.078 & 0.909 \\
ULAS J111429.54+072809.5 & \citetalias{prime4} & esdL0  & $-$1.4  & This paper & 0.129 & 0.276 & 0.124 & 0.877	\\
SDSS J124410.11+273625.8 &\citealt{lodi12} & esdL0.5 & $-$1.5 & \citetalias{prime1} & 0.163 & 0.284 & 0.138 & 0.877 \\
ULAS J135216.31+312327.0 & \citetalias{prime4} & esdL0.5  & $-$1.6  & This paper & 0.130 & 0.270 & 0.119 & 0.880	\\
ULAS J020858.62+020657.0 & \citetalias{prime3}  & esdL3 & $-$1.5 & \citetalias{prime3} & 0.098 & 0.141 & 0.267 & 0.704 \\
SDSS J084648.88+302801.7 &\citealt{zha13} & usdM6 & $-$2.0 & This paper & 0.308 & 0.431 & 0.897 & 0.122 \\
APMPM J0559$-$2903 &\citealt{schw99} & usdM7 & $-$1.8 & \citetalias{prime1} & 0.217 & 0.331 & 0.604 & 0.421 \\	
LEHPM 2-59 & \citetalias{burg06}  & usdM8 & $-$2.2 & \citetalias{prime1} & 0.175 & 0.265 & 0.656 & 0.349 \\
SSSPM J10130734$-$1356204 & \citealt{scho04} & usdL0  & $-$1.8 & \citetalias{prime1} & 0.114 & 0.204 & 0.248 & 0.734 \\
SDSS J010448.46+153501.8 & \citealt{lodi12} & usdL1.5 & $-$2.4 & \citetalias{prime2} & 0.122 & 0.133 & 0.356 & 0.620 \\
SDSS J125637.16$-$022452.2 & \citealt{siva09} & usdL3 & $-$1.8 & \citetalias{prime1} & 0.103 & 0.118 & 0.188 & 0.777 \\
2MASS J16262034+3925190 & \citealt{burg04b} & usdL4  & $-$1.8 & \citetalias{prime1} & 0.098 & 0.131 & 0.260 & 0.709 \\
ULAS J230711.01+014447.1 & \citetalias{prime3} & usdL4.5 & $-$1.7 & \citetalias{prime3} & 0.069 & 0.145 & 0.282 & 0.687 \\
\hline
\end{tabular}
\end{table*}

M subclasses are defined by ranges of the $\zeta_{\rm TiO/CaH}$ index. M dwarfs have a mean value of $\zeta_{\rm TiO/CaH} = 1$. The boundaries between dM, sdM, esdM, and usdM subclasses are located at $\zeta_{\rm CaH/TiO}$ = 0.825, 0.5, and 0.2, respectively. Fig. \ref{fsmz} (b) shows that the $\zeta_{\rm TiO/CaH}$ index is correlated to [Fe/H] for M0--3 subdwarfs, and can be fitted with a polynomial function described as:
\begin{equation}
\label{equ1}
{\rm [Fe/H]} = 1.949 \zeta_{\rm TiO/CaH} - 1.853
\end{equation}
with a root mean square (rms) of 0.252. However, the $\zeta_{\rm TiO/CaH}$ index is not correlated to [Fe/H] for late-type M and early-type L subdwarfs. The two metallicity trace lines jump off the $\zeta_{\rm CaH/TiO}$--[Fe/H] correlation (Equation \ref{equ1}) at esdM5.5 and usdM6 types, and move further away at late-type M and L types. 

Fig. \ref{fsmz} (c) shows the variation of the $\zeta_{\rm TiO/CaH}$ index following the two metallicity trace lines across M and L spectral types. The $\zeta_{\rm CaH/TiO}$ index is clearly not consistent across mid- and late-type M subdwarfs. Consequently, M6+ subdwarfs in the metallicity range as esdM0--5 subdwarfs would be classified into sdM and dM subclasses, and M6.5+ subdwarfs in the metallicity range as usdM0--5 subdwarfs would be classified into esdM and sdM subclasses by the $\zeta_{\rm TiO/CaH}$ index. Likewise, some late-type M subdwarfs in the metallicity range as sdM0--5 subdwarfs would be classified into dM subclass by the $\zeta_{\rm TiO/CaH}$ index.  

Fig. \ref{fsmz} (d) shows the two metallicity trace lines presented in Fig. \ref{fsmz} (a) and boundaries between different subclasses in a parameter space of TiO5, CaH2, and CaH3 indices. These two metallicity trace lines follow the boundary defined by the $\zeta_{\rm CaH/TiO}$ index for early- and mid-type M subdwarfs. However, the esd and usd trace lines turned to the left and crossed the M subclass boundaries after esdM5.5 and usdM6 types, respectively. Fig. \ref{ftiob} shows the same problem exist in other M subdwarf classification schemes based on CaH and TiO spectral indices \citep{gizi97,dhit12,zhan19}. Figs \ref{fsmz} (b-d) and \ref{ftiob} show that the  $\zeta_{\rm CaH/TiO}$ index and the TiO and CaH spectral indices are temperature dependent after esdM5.5 and usdM6 types, corresponding to a mass around 0.1 M$_{\sun}$ \citepalias[fig. 9;][]{prime2}. Figs \ref{fsmz} and \ref{ftiob} show that the metallicity of each M subclass is inconsistent across mid- to late-types. 

Although, the classification scheme of M subdwarfs were slightly refined a few times \citep{dhit12,lepi13,zhan19} after \citet{lepi07}. However, the $\zeta_{\rm TiO/CaH}$ index have small updates mostly affecting early-type M subdwarfs, and the metallicity of each subclass is still not consistent across all M subtypes. 
This is because UCSDs have significant different atmospheres from early-type M subdwarfs. First, dust starts to form in the ultracool atmospheres of late-type M dwarfs/subdwarfs, and have significant impact on spectral indices of UCDs/UCSDs \citep{jone97,burr99}. Secondly, late-type M and L subdwarfs have higher gravity than early-type M subdwarfs. Most of available oxygen would form H$_2$O in metal-poor atmospheres under high pressure, thus TiO reduces more rapidly with decreasing metallicity in the atmospheres of UCSDs \citep[e.g.,][Section 4.6.4]{reid05}. However, this does not affect hydride absorptions (CaH, and FeH). Therefore, the metallicity index $\zeta_{\rm TiO/CaH}$ should not be used to classify late-type M and L subdwarfs.

\section{Mass ranges of M subclasses}
\label{sms}
Hydrodynamical simulations of the formation of stars and brown dwarfs in star clusters show that there is no significant dependence of stellar properties (initial mass function, binary fraction) on opacity or metallicity at 0.01--4 M$_{\sun}$ and 0.01--3 $Z_{\sun}$ \citep{bate14,bate19}. However, the lack of late-type M extreme and ultra subdwarfs has been noticed \citep{mone92,gizi97}. \citet[table 9]{zha13} show that the binary fraction of M subdwarfs is decreasing from the sdM to the esdM and usdM subclasses under on classification scheme of \citet{lepi07}. These can be explained by the special properties of M subdwarfs due to low opacity and the biased spectral classification of M subdwarfs.

First, M subdwarfs under current classification scheme have lower mass ranges than M dwarfs \citepalias[fig. 9;][]{prime2}. The maximum temperatures to form TiO, CaH, and H$_2$O molecules are lower at lower metallicity \citep[e.g. fig. 9;][]{jao08}. At a certain mass, very low-mass stars with lower metallicity would have higher $T_{\rm eff}$. Consequently, depending on the metallicity, an M0 subdwarf could have a mass up to about five times lower than a field M0 dwarf. The masses of usdM0, esdM0, sdM0 subdwarfs, and M0 dwarfs are around 0.13--0.15, 0.15--0.3, 0.3--0.5, and 0.5--0.6 M$_{\sun}$, respectively. This is why M subdwarfs have much smaller radii than M dwarfs with the same $T_{\rm eff}$ \citep[fig. 9;][]{kess19}.  For example, the cool subdwarf component of the usdK7+WD eclipsing binary (SDSS J235524.29+044855.7) has mass of 0.1502$\pm$0.0017 M$_{\sun}$, radius of 0.1821$\pm$0.0007 R$_{\sun}$, $T_{\rm eff} = 3650\pm50$ K, and [Fe/H] = $-1.55\pm0.25$ \citep{reba19}. Secondly, metallicity scale of late-type M subdwarfs are overestimated by spectral indices in current classification schemes due to higher gravity and dust formation in ultracool atmospheres. Consequently, late-type M subdwarfs in the matallicity ranges of early-type sdM/esdM/usdM are classified into dM/sdM/esdM subclasses, respectively (see Section \ref{smc}).

The spectral type classified by broad absorption bands \citep[e.g. CaH, TiO;][]{gizi97,lepi07} in observed spectra of M subdwarfs is mainly related to their $T_{\rm eff}$ and metallicity. Since red subdwarfs have subsolar abundance. M subdwarfs are hotter than M dwarfs with equivalent mass (\citealt{chab97,burr01}; \citetalias{prime2}). The CaH and TiO molecular bands (identification of red dwarf/subdwarfs) appear in atmospheres of red subdwarfs at lower temperature than in red dwarfs. Consequently, M subdwarfs have much lower and narrower mass range than M dwarfs. 

L subdwafs have much narrower mass range than L dwarfs, because the corresponding spectral types of the substellar transition zone among UCSDs is from early-L to mid-T types. The spectral type (and $T_{\rm eff}$) sampling from early-type L to mid-type T subdwarfs are stretched by the substellar transition zone \citep{zha18c} which covers a narrow mass range. The mass range of the substellar transition zone is between 0.065 and 0.079 M$_{\sun}$ for solar metallicity, and slightly higher but narrower at lower metallicities \citepalias[fig. 5;][]{prime6}. 

\citet[fig.10]{zhan19} show that the fraction of active M subdwarfs is smaller and peaks at earlier subtypes than that of M dwarfs. This is partially due to the fact that M subdwarfs have lower mass ranges than M dwarfs. The comparison between M dwarfs and M subdwarfs is between populations in different mass ranges.

\section{Summary and conclusions}
\label{scon}
I presented the discovery of the first wide M + L extreme subdwarf binary (Gaia J0452$-$36AB) at a distance of 137.27$^{+0.68}_{-0.67}$ pc. The binary has typical metallicity ([Fe/H] $= -1.4\pm0.2$) and space velocity ($U = 65\pm$9 km s$^{-1}$; $V = -129\pm$15 km s$^{-1}$; and $W = 42\pm$14 km s$^{-1}$) for halo population. Gaia J0452$-$36AB is an esdM1+esdL0 subdwarf pair with $T_{\rm eff}$ of 3550$\pm$100 and 2600$\pm$100 K, respectively. Gaia J0452$-$36AB have masses of 0.151$^{+0.029}_{-0.019}$ and 0.0855$^{+0.0014}_{-0.0010}$ M$_{\sun}$, respectively. Both components are very low-mass stars above the substellar transition zone at their metallicity. The binary has a projected separation of 15828$\pm$78 au, corresponding to 0.092 $r_{\rm J}$, and is a gravitationally bound system.

I also presented the discovery of Ruiz 440-469B, a wide M8 dwarf companion to a cool DA white dwarf, Ruiz 440-469A. Ruiz 440-469B has slightly bluer $J-K$ colour than main-sequence stars thus maybe mildly metal-poor. Ruiz 440-469AB is at a distance of 112.94$^{+3.77}_{-3.53}$ pc with a projected separation of 8794$^{+292}_{-174}$ au and a tangential velocity of 112.91$^{+3.77}_{-3.53}$ km s$^{-1}$. Gaia J0452$-$36AB is also a gravitationally bound system. 

Late-type M and L subdwarfs have high gravity and dust in their atmospheres which have significant impacts on the presences of CaH and TiO absorption bands. The $\zeta_{\rm CaH/TiO}$ index is invalid for the classification of late-type M and L subdwarfs.  
As shown in Figs \ref{fsmz} and \ref{ftiob}, the existing classification schemes of M subdwarfs performed well for M0--5 subdwarfs, but did not pass the metallicity consistency test for late-type M subdwarfs. Inconsistent metallicity across subtypes in each subclass would provide misleading information on the metallicity scale of late-type M subdwarfs. 

M subdwarfs have lower and narrower mass range than M dwarfs. Because the cool subdwarfs have higher $T_{\rm eff}$ than dwarfs with equivalent mass, but the molecules used to define red dwarf/subdwarfs or M spectral class form at lower temperature in atmospheres with lower metallicity. The comparison between M dwarfs and M subdwarfs is between populations in different mass ranges. This partially explains the lack of late-type M extreme and ultra subdwarfs and decreasing binary fraction from sdM, to esdM and usdM subclass, in addition to the biased classification of M subdwarfs. 

Red dwarfs have low masses and small radii, thus signals of RV variation and transit due to orbiting planets are relatively significant. Nearby red dwarfs became popular targets for searches of rocky exoplanets of the habitable zone. Red subdwarfs of the thick disc are in the the Galactic habitable zone \citep{line04} and are also good targets for exoplanet searches \citep[e.g.,][]{angl14}. 
Mid- to late-type K subdwarfs are also in the category of very low-mass stars with mass $\la$ 0.5 M$_{\sun}$ and radius $\la$ 0.5 R$_{\sun}$, thus are also good targets for exoplanet searches with RV and transit monitoring, in addition to M subdwarfs. 

\section*{Acknowledgements}
This article is based on observations made in the Observatorios de Canarias del IAC with the William Herschel Telescope (WHT) operated on the island of La Palma by the Isaac Newton Group of Telescopes in the Observatorio del los Muchachos. Based on observations collected at the European Organisation for Astronomical Research in the Southern Hemisphere under ESO programmes 096.C-0130 and 098.D-0222. 
This work presents results from the European Space Agency (ESA) space mission {\sl Gaia}. {\sl Gaia} data is being processed by the {\sl Gaia} Data Processing and Analysis Consortium (DPAC). Funding for the DPAC is provided by national institutions, in particular the institutions participating in the {\sl Gaia} MultiLateral Agreement (MLA). The {\sl Gaia} mission website is \url{https://www.cosmos.esa.int/gaia}. The {\sl Gaia} archive website is \url{https://archives.esac.esa.int/gaia}. 
Based on observations obtained as part of the VISTA Hemisphere Survey, ESO Progam, 179.A-2010 (PI: McMahon). Data processing has been contributed by the VISTA Data Flow System at CASU, Cambridge and WFAU, Edinburgh. The VISTA Data Flow System pipeline processing and science archive are described in \citet{irwi04}, \citet{hamb08} and \citet{cros12}. 
This publication makes use of data products from the {\sl Wide-field Infrared Survey Explorer}, which is a joint project of the University of California, Los Angeles, and the Jet Propulsion Laboratory/California Institute of Technology, funded by the National Aeronautics and Space Administration. 
Synthetic spectra used in this paper are calculated by France Allard and Derek Homeier based on BT-Dusty model atmospheres developed by France Allard. 
ZHZ was supported by the PSL fellowship.

\bibliographystyle{mnras}
\bibliography{primeval7} 

\begin{thebibliography}{}
\makeatletter
\relax
\def\mn@urlcharsother{\let\do\@makeother \do\$\do\&\do\#\do\^\do\_\do\%\do\~}
\def\mn@doi{\begingroup\mn@urlcharsother \@ifnextchar [ {\mn@doi@}
  {\mn@doi@[]}}
\def\mn@doi@[#1]#2{\def\@tempa{#1}\ifx\@tempa\@empty \href
  {http://dx.doi.org/#2} {doi:#2}\else \href {http://dx.doi.org/#2} {#1}\fi
  \endgroup}
\def\mn@eprint#1#2{\mn@eprint@#1:#2::\@nil}
\def\mn@eprint@arXiv#1{\href {http://arxiv.org/abs/#1} {{\tt arXiv:#1}}}
\def\mn@eprint@dblp#1{\href {http://dblp.uni-trier.de/rec/bibtex/#1.xml}
  {dblp:#1}}
\def\mn@eprint@#1:#2:#3:#4\@nil{\def\@tempa {#1}\def\@tempb {#2}\def\@tempc
  {#3}\ifx \@tempc \@empty \let \@tempc \@tempb \let \@tempb \@tempa \fi \ifx
  \@tempb \@empty \def\@tempb {arXiv}\fi \@ifundefined
  {mn@eprint@\@tempb}{\@tempb:\@tempc}{\expandafter \expandafter \csname
  mn@eprint@\@tempb\endcsname \expandafter{\@tempc}}}

\bibitem[\protect\citeauthoryear{{Aganze} et~al.,}{{Aganze}
  et~al.}{2016}]{agna16}
{Aganze} C.,  et~al., 2016, \mn@doi [\aj] {10.3847/0004-6256/151/2/46}, \href
  {https://ui.adsabs.harvard.edu/abs/2016AJ....151...46A} {151, 46}

\bibitem[\protect\citeauthoryear{{Allard}}{{Allard}}{2014}]{alla14}
{Allard} F.,  2014, in {Booth} M.,  {Matthews} B.~C.,   {Graham} J.~R.,  eds,
  IAU Symposium Vol. 299, Exploring the Formation and Evolution of Planetary
  Systems. pp 271--272, \mn@doi{10.1017/S1743921313008545}

\bibitem[\protect\citeauthoryear{{Allard} \& {Hauschildt}}{{Allard} \&
  {Hauschildt}}{1995}]{alla95}
{Allard} F.,  {Hauschildt} P.~H.,  1995, \mn@doi [\apj] {10.1086/175708}, \href
  {https://ui.adsabs.harvard.edu/abs/1995ApJ...445..433A} {445, 433}

\bibitem[\protect\citeauthoryear{{Anglada-Escude} et~al.,}{{Anglada-Escude}
  et~al.}{2014}]{angl14}
{Anglada-Escude} G.,  et~al., 2014, \mn@doi [\mnras] {10.1093/mnrasl/slu076},
  \href {https://ui.adsabs.harvard.edu/abs/2014MNRAS.443L..89A} {443, L89}

\bibitem[\protect\citeauthoryear{{Baraffe} \& {Chabrier}}{{Baraffe} \&
  {Chabrier}}{2018}]{bara18}
{Baraffe} I.,  {Chabrier} G.,  2018, \mn@doi [\aap]
  {10.1051/0004-6361/201834062}, \href
  {http://adsabs.harvard.edu/abs/2018A%26A...619A.177B} {619, A177}

\bibitem[\protect\citeauthoryear{{Baraffe}, {Chabrier}, {Allard}  \&
  {Hauschildt}}{{Baraffe} et~al.}{1997}]{bara97}
{Baraffe} I.,  {Chabrier} G.,  {Allard} F.,   {Hauschildt} P.~H.,  1997, \aap,
  \href {https://ui.adsabs.harvard.edu/abs/1997A&A...327.1054B} {327, 1054}

\bibitem[\protect\citeauthoryear{{Bate}}{{Bate}}{2014}]{bate14}
{Bate} M.~R.,  2014, \mn@doi [\mnras] {10.1093/mnras/stu795}, \href
  {https://ui.adsabs.harvard.edu/abs/2014MNRAS.442..285B} {442, 285}

\bibitem[\protect\citeauthoryear{{Bate}}{{Bate}}{2019}]{bate19}
{Bate} M.~R.,  2019, \mn@doi [\mnras] {10.1093/mnras/stz103}, \href
  {https://ui.adsabs.harvard.edu/abs/2019MNRAS.484.2341B} {484, 2341}

\bibitem[\protect\citeauthoryear{{Benn}, {Dee}  \& {Ag{\'o}cs}}{{Benn}
  et~al.}{2008}]{benn08}
{Benn} C.,  {Dee} K.,   {Ag{\'o}cs} T.,  2008, in Ground-based and Airborne
  Instrumentation for Astronomy II. p. 70146X, \mn@doi{10.1117/12.788694}

\bibitem[\protect\citeauthoryear{{Bowler}, {Liu}  \& {Cushing}}{{Bowler}
  et~al.}{2009}]{bowl09}
{Bowler} B.~P.,  {Liu} M.~C.,   {Cushing} M.~C.,  2009, \mn@doi [\apj]
  {10.1088/0004-637X/706/2/1114}, \href
  {https://ui.adsabs.harvard.edu/abs/2009ApJ...706.1114B} {706, 1114}

\bibitem[\protect\citeauthoryear{{Burgasser}}{{Burgasser}}{2004}]{burg04b}
{Burgasser} A.~J.,  2004, \mn@doi [\apjl] {10.1086/425418}, \href
  {http://adsabs.harvard.edu/abs/2004ApJ...614L..73B} {614, L73}

\bibitem[\protect\citeauthoryear{{Burgasser} \& {Kirkpatrick}}{{Burgasser} \&
  {Kirkpatrick}}{2006}]{burg06}
{Burgasser} A.~J.,  {Kirkpatrick} J.~D.,  2006, \mn@doi [\apj]
  {10.1086/504375}, \href {http://adsabs.harvard.edu/abs/2006ApJ...645.1485B}
  {645, 1485}

\bibitem[\protect\citeauthoryear{{Burgasser} et~al.,}{{Burgasser}
  et~al.}{2002}]{burg02}
{Burgasser} A.~J.,  et~al., 2002, \mn@doi [\apj] {10.1086/324033}, \href
  {http://adsabs.harvard.edu/abs/2002ApJ...564..421B} {564, 421}

\bibitem[\protect\citeauthoryear{{Burgasser} et~al.,}{{Burgasser}
  et~al.}{2003}]{burg03}
{Burgasser} A.~J.,  et~al., 2003, \mn@doi [\apj] {10.1086/375813}, \href
  {https://ui.adsabs.harvard.edu/abs/2003ApJ...592.1186B} {592, 1186}

\bibitem[\protect\citeauthoryear{{Burgasser}, {McElwain}, {Kirkpatrick},
  {Cruz}, {Tinney}  \& {Reid}}{{Burgasser} et~al.}{2004}]{burg04}
{Burgasser} A.~J.,  {McElwain} M.~W.,  {Kirkpatrick} J.~D.,  {Cruz} K.~L.,
  {Tinney} C.~G.,   {Reid} I.~N.,  2004, \mn@doi [\aj] {10.1086/383549}, \href
  {http://adsabs.harvard.edu/abs/2004AJ....127.2856B} {127, 2856}

\bibitem[\protect\citeauthoryear{{Burgasser}, {Cruz}  \&
  {Kirkpatrick}}{{Burgasser} et~al.}{2007}]{burg07}
{Burgasser} A.~J.,  {Cruz} K.~L.,   {Kirkpatrick} J.~D.,  2007, \mn@doi [\apj]
  {10.1086/510148}, \href {http://adsabs.harvard.edu/abs/2007ApJ...657..494B}
  {657, 494}

\bibitem[\protect\citeauthoryear{{Burningham} et~al.,}{{Burningham}
  et~al.}{2010}]{burn10}
{Burningham} B.,  et~al., 2010, \mn@doi [\mnras]
  {10.1111/j.1365-2966.2010.16411.x}, \href
  {https://ui.adsabs.harvard.edu/abs/2010MNRAS.404.1952B} {404, 1952}

\bibitem[\protect\citeauthoryear{{Burningham}, {Smith}, {Cardoso}, {Lucas},
  {Burgasser}, {Jones}  \& {Smart}}{{Burningham} et~al.}{2014}]{burn14}
{Burningham} B.,  {Smith} L.,  {Cardoso} C.~V.,  {Lucas} P.~W.,  {Burgasser}
  A.~J.,  {Jones} H.~R.~A.,   {Smart} R.~L.,  2014, \mn@doi [\mnras]
  {10.1093/mnras/stu184}, \href
  {https://ui.adsabs.harvard.edu/abs/2014MNRAS.440..359B} {440, 359}

\bibitem[\protect\citeauthoryear{{Burrows} \& {Sharp}}{{Burrows} \&
  {Sharp}}{1999}]{burr99}
{Burrows} A.,  {Sharp} C.~M.,  1999, \mn@doi [\apj] {10.1086/306811}, \href
  {http://adsabs.harvard.edu/abs/1999ApJ...512..843B} {512, 843}

\bibitem[\protect\citeauthoryear{{Burrows}, {Hubbard}, {Lunine}  \&
  {Liebert}}{{Burrows} et~al.}{2001}]{burr01}
{Burrows} A.,  {Hubbard} W.~B.,  {Lunine} J.~I.,   {Liebert} J.,  2001, \mn@doi
  [Reviews of Modern Physics] {10.1103/RevModPhys.73.719}, \href
  {https://ui.adsabs.harvard.edu/abs/2001RvMP...73..719B} {73, 719}

\bibitem[\protect\citeauthoryear{{Chabrier} \& {Baraffe}}{{Chabrier} \&
  {Baraffe}}{1997}]{chab97}
{Chabrier} G.,  {Baraffe} I.,  1997, \aap, \href
  {https://ui.adsabs.harvard.edu/abs/1997A&A...327.1039C} {327, 1039}

\bibitem[\protect\citeauthoryear{{Cross} et~al.,}{{Cross}
  et~al.}{2012}]{cros12}
{Cross} N.~J.~G.,  et~al., 2012, \mn@doi [\aap] {10.1051/0004-6361/201219505},
  \href {http://adsabs.harvard.edu/abs/2012A%26A...548A.119C} {548, A119}

\bibitem[\protect\citeauthoryear{{Dhital}, {West}, {Stassun}, {Bochanski},
  {Massey}  \& {Bastien}}{{Dhital} et~al.}{2012}]{dhit12}
{Dhital} S.,  {West} A.~A.,  {Stassun} K.~G.,  {Bochanski} J.~J.,  {Massey}
  A.~P.,   {Bastien} F.~A.,  2012, \mn@doi [\aj] {10.1088/0004-6256/143/3/67},
  \href {http://adsabs.harvard.edu/abs/2012AJ....143...67D} {143, 67}

\bibitem[\protect\citeauthoryear{{Freudling}, {Romaniello}, {Bramich},
  {Ballester}, {Forchi}, {Garc{\'{\i}}a-Dabl{\'o}}, {Moehler}  \&
  {Neeser}}{{Freudling} et~al.}{2013}]{freu13}
{Freudling} W.,  {Romaniello} M.,  {Bramich} D.~M.,  {Ballester} P.,  {Forchi}
  V.,  {Garc{\'{\i}}a-Dabl{\'o}} C.~E.,  {Moehler} S.,   {Neeser} M.~J.,  2013,
  \mn@doi [\aap] {10.1051/0004-6361/201322494}, \href
  {http://adsabs.harvard.edu/abs/2013A%26A...559A..96F} {559, A96}

\bibitem[\protect\citeauthoryear{{Gaia Collaboration} et~al.,}{{Gaia
  Collaboration} et~al.}{2018}]{gaia18}
{Gaia Collaboration} et~al., 2018, \mn@doi [\aap]
  {10.1051/0004-6361/201833051}, \href
  {http://adsabs.harvard.edu/abs/2018A%26A...616A...1G} {616, A1}

\bibitem[\protect\citeauthoryear{{Gill}}{{Gill}}{1899}]{gill99}
{Gill} D.,  1899, The Observatory, \href
  {https://ui.adsabs.harvard.edu/abs/1899Obs....22...99G} {22, 99}

\bibitem[\protect\citeauthoryear{{Gizis}}{{Gizis}}{1997}]{gizi97}
{Gizis} J.~E.,  1997, \mn@doi [\aj] {10.1086/118302}, \href
  {http://adsabs.harvard.edu/abs/1997AJ....113..806G} {113, 806}

\bibitem[\protect\citeauthoryear{{Hambly} et~al.,}{{Hambly}
  et~al.}{2008}]{hamb08}
{Hambly} N.~C.,  et~al., 2008, \mn@doi [\mnras]
  {10.1111/j.1365-2966.2007.12700.x}, \href
  {http://adsabs.harvard.edu/abs/2008MNRAS.384..637H} {384, 637}

\bibitem[\protect\citeauthoryear{{Hertzsprung}}{{Hertzsprung}}{1909}]{hert09}
{Hertzsprung} E.,  1909, \mn@doi [Astronomische Nachrichten]
  {10.1002/asna.19081792402}, \href
  {https://ui.adsabs.harvard.edu/abs/1909AN....179..373H} {179, 373}

\bibitem[\protect\citeauthoryear{{Irwin} et~al.,}{{Irwin}
  et~al.}{2004}]{irwi04}
{Irwin} M.~J.,  et~al., 2004, in {Quinn} P.~J.,  {Bridger} A.,  eds,  \procspie
  Vol. 5493, Optimizing Scientific Return for Astronomy through Information
  Technologies. pp 411--422, \mn@doi{10.1117/12.551449}

\bibitem[\protect\citeauthoryear{{Jao}, {Henry}, {Beaulieu}  \&
  {Subasavage}}{{Jao} et~al.}{2008}]{jao08}
{Jao} W.-C.,  {Henry} T.~J.,  {Beaulieu} T.~D.,   {Subasavage} J.~P.,  2008,
  \mn@doi [\aj] {10.1088/0004-6256/136/2/840}, \href
  {https://ui.adsabs.harvard.edu/abs/2008AJ....136..840J} {136, 840}

\bibitem[\protect\citeauthoryear{{Jao}, {Henry}, {Winters}, {Subasavage},
  {Riedel}, {Silverstein}  \& {Ianna}}{{Jao} et~al.}{2017}]{jao17}
{Jao} W.-C.,  {Henry} T.~J.,  {Winters} J.~G.,  {Subasavage} J.~P.,  {Riedel}
  A.~R.,  {Silverstein} M.~L.,   {Ianna} P.~A.,  2017, \mn@doi [\aj]
  {10.3847/1538-3881/aa8b64}, \href
  {https://ui.adsabs.harvard.edu/abs/2017AJ....154..191J} {154, 191}

\bibitem[\protect\citeauthoryear{{Jao}, {Henry}, {Gies}  \& {Hambly}}{{Jao}
  et~al.}{2018}]{jao18}
{Jao} W.-C.,  {Henry} T.~J.,  {Gies} D.~R.,   {Hambly} N.~C.,  2018, \mn@doi
  [\apjl] {10.3847/2041-8213/aacdf6}, \href
  {http://adsabs.harvard.edu/abs/2018ApJ...861L..11J} {861, L11}

\bibitem[\protect\citeauthoryear{{Jiang} \& {Tremaine}}{{Jiang} \&
  {Tremaine}}{2010}]{jian10}
{Jiang} Y.-F.,  {Tremaine} S.,  2010, \mn@doi [\mnras]
  {10.1111/j.1365-2966.2009.15744.x}, \href
  {https://ui.adsabs.harvard.edu/abs/2010MNRAS.401..977J} {401, 977}

\bibitem[\protect\citeauthoryear{{Jones} \& {Tsuji}}{{Jones} \&
  {Tsuji}}{1997}]{jone97}
{Jones} H.~R.~A.,  {Tsuji} T.,  1997, \mn@doi [\apjl] {10.1086/310619}, \href
  {http://adsabs.harvard.edu/abs/1997ApJ...480L..39J} {480, L39}

\bibitem[\protect\citeauthoryear{{Kapteyn}}{{Kapteyn}}{1897}]{kapt97}
{Kapteyn} J.~C.,  1897, \mn@doi [Astronomische Nachrichten]
  {10.1002/asna.18981450906}, \href
  {https://ui.adsabs.harvard.edu/abs/1897AN....145..159K} {145, 159}

\bibitem[\protect\citeauthoryear{{Kesseli} et~al.,}{{Kesseli}
  et~al.}{2019}]{kess19}
{Kesseli} A.~Y.,  et~al., 2019, \mn@doi [\aj] {10.3847/1538-3881/aae982}, \href
  {http://adsabs.harvard.edu/abs/2019AJ....157...63K} {157, 63}

\bibitem[\protect\citeauthoryear{{Kirkpatrick}, {Henry}  \&
  {Irwin}}{{Kirkpatrick} et~al.}{1997}]{kirk97}
{Kirkpatrick} J.~D.,  {Henry} T.~J.,   {Irwin} M.~J.,  1997, \mn@doi [\aj]
  {10.1086/118357}, \href
  {https://ui.adsabs.harvard.edu/abs/1997AJ....113.1421K} {113, 1421}

\bibitem[\protect\citeauthoryear{{Kirkpatrick} et~al.,}{{Kirkpatrick}
  et~al.}{1999}]{kirk99}
{Kirkpatrick} J.~D.,  et~al., 1999, \mn@doi [\apj] {10.1086/307414}, \href
  {http://adsabs.harvard.edu/abs/1999ApJ...519..802K} {519, 802}

\bibitem[\protect\citeauthoryear{{Kirkpatrick} et~al.,}{{Kirkpatrick}
  et~al.}{2014}]{kirk14}
{Kirkpatrick} J.~D.,  et~al., 2014, \mn@doi [\apj]
  {10.1088/0004-637X/783/2/122}, \href
  {http://adsabs.harvard.edu/abs/2014ApJ...783..122K} {783, 122}

\bibitem[\protect\citeauthoryear{{Kirkpatrick} et~al.,}{{Kirkpatrick}
  et~al.}{2016}]{kirk16}
{Kirkpatrick} J.~D.,  et~al., 2016, \mn@doi [\apjs]
  {10.3847/0067-0049/224/2/36}, \href
  {http://adsabs.harvard.edu/abs/2016ApJS..224...36K} {224, 36}

\bibitem[\protect\citeauthoryear{{Koren}, {Blake}, {Dahn}  \& {Harris}}{{Koren}
  et~al.}{2016}]{kore16}
{Koren} S.~C.,  {Blake} C.~H.,  {Dahn} C.~C.,   {Harris} H.~C.,  2016, \mn@doi
  [\aj] {10.3847/0004-6256/151/3/57}, \href
  {http://adsabs.harvard.edu/abs/2016AJ....151...57K} {151, 57}

\bibitem[\protect\citeauthoryear{{Kotoneva}, {Innanen}, {Dawson}, {Wood}  \&
  {De Robertis}}{{Kotoneva} et~al.}{2005}]{koto05}
{Kotoneva} E.,  {Innanen} K.,  {Dawson} P.~C.,  {Wood} P.~R.,   {De Robertis}
  M.~M.,  2005, \mn@doi [\aap] {10.1051/0004-6361:20042287}, \href
  {https://ui.adsabs.harvard.edu/abs/2005A&A...438..957K} {438, 957}

\bibitem[\protect\citeauthoryear{{Kuiper}}{{Kuiper}}{1939}]{kuip39}
{Kuiper} G.~P.,  1939, \mn@doi [\apj] {10.1086/144075}, \href
  {https://ui.adsabs.harvard.edu/abs/1939ApJ....89..548K} {89, 548}

\bibitem[\protect\citeauthoryear{{Kuiper}}{{Kuiper}}{1940}]{kuip40}
{Kuiper} G.~P.,  1940, \mn@doi [\apj] {10.1086/144166}, \href
  {https://ui.adsabs.harvard.edu/abs/1940ApJ....91..269K} {91, 269}

\bibitem[\protect\citeauthoryear{{Lawrence} et~al.,}{{Lawrence}
  et~al.}{2007}]{lawr07}
{Lawrence} A.,  et~al., 2007, \mn@doi [\mnras]
  {10.1111/j.1365-2966.2007.12040.x}, \href
  {https://ui.adsabs.harvard.edu/abs/2007MNRAS.379.1599L} {379, 1599}

\bibitem[\protect\citeauthoryear{{L{\'e}pine} \& {Scholz}}{{L{\'e}pine} \&
  {Scholz}}{2008}]{lepi08}
{L{\'e}pine} S.,  {Scholz} R.-D.,  2008, \mn@doi [\apj] {10.1086/590183}, \href
  {https://ui.adsabs.harvard.edu/abs/2008ApJ...681L..33L} {681, L33}

\bibitem[\protect\citeauthoryear{{L{\'e}pine}, {Shara}  \& {Rich}}{{L{\'e}pine}
  et~al.}{2003a}]{lepi03a}
{L{\'e}pine} S.,  {Shara} M.~M.,   {Rich} R.~M.,  2003a, \mn@doi [\apj]
  {10.1086/374210}, \href
  {https://ui.adsabs.harvard.edu/abs/2003ApJ...585L..69L} {585, L69}

\bibitem[\protect\citeauthoryear{{L{\'e}pine}, {Rich}  \& {Shara}}{{L{\'e}pine}
  et~al.}{2003b}]{lepi03}
{L{\'e}pine} S.,  {Rich} R.~M.,   {Shara} M.~M.,  2003b, \mn@doi [\apjl]
  {10.1086/377069}, \href {http://adsabs.harvard.edu/abs/2003ApJ...591L..49L}
  {591, L49}

\bibitem[\protect\citeauthoryear{{L{\'e}pine}, {Rich}  \& {Shara}}{{L{\'e}pine}
  et~al.}{2007}]{lepi07}
{L{\'e}pine} S.,  {Rich} R.~M.,   {Shara} M.~M.,  2007, \mn@doi [\apj]
  {10.1086/521614}, \href {http://adsabs.harvard.edu/abs/2007ApJ...669.1235L}
  {669, 1235}

\bibitem[\protect\citeauthoryear{{L{\'e}pine}, {Hilton}, {Mann}, {Wilde},
  {Rojas-Ayala}, {Cruz}  \& {Gaidos}}{{L{\'e}pine} et~al.}{2013}]{lepi13}
{L{\'e}pine} S.,  {Hilton} E.~J.,  {Mann} A.~W.,  {Wilde} M.,  {Rojas-Ayala}
  B.,  {Cruz} K.~L.,   {Gaidos} E.,  2013, \mn@doi [\aj]
  {10.1088/0004-6256/145/4/102}, \href
  {http://adsabs.harvard.edu/abs/2013AJ....145..102L} {145, 102}

\bibitem[\protect\citeauthoryear{{Lineweaver}, {Fenner}  \&
  {Gibson}}{{Lineweaver} et~al.}{2004}]{line04}
{Lineweaver} C.~H.,  {Fenner} Y.,   {Gibson} B.~K.,  2004, \mn@doi [Science]
  {10.1126/science.1092322}, \href
  {https://ui.adsabs.harvard.edu/abs/2004Sci...303...59L} {303, 59}

\bibitem[\protect\citeauthoryear{{Lodieu}, {Espinoza Contreras}, {Zapatero
  Osorio}, {Solano}, {Aberasturi}  \& {Mart{\'{\i}}n}}{{Lodieu}
  et~al.}{2012}]{lodi12}
{Lodieu} N.,  {Espinoza Contreras} M.,  {Zapatero Osorio} M.~R.,  {Solano} E.,
  {Aberasturi} M.,   {Mart{\'{\i}}n} E.~L.,  2012, \mn@doi [\aap]
  {10.1051/0004-6361/201118717}, \href
  {http://adsabs.harvard.edu/abs/2012A%26A...542A.105L} {542, A105}

\bibitem[\protect\citeauthoryear{{Lodieu}, {Espinoza Contreras}, {Zapatero
  Osorio}, {Solano}, {Aberasturi}, {Mart{\'\i}n}  \& {Rodrigo}}{{Lodieu}
  et~al.}{2017}]{lodi17}
{Lodieu} N.,  {Espinoza Contreras} M.,  {Zapatero Osorio} M.~R.,  {Solano} E.,
  {Aberasturi} M.,  {Mart{\'\i}n} E.~L.,   {Rodrigo} C.,  2017, \mn@doi [\aap]
  {10.1051/0004-6361/201629410}, \href
  {https://ui.adsabs.harvard.edu/abs/2017A&A...598A..92L} {598, A92}

\bibitem[\protect\citeauthoryear{{Luhman} \& {Sheppard}}{{Luhman} \&
  {Sheppard}}{2014}]{luhm14}
{Luhman} K.~L.,  {Sheppard} S.~S.,  2014, \mn@doi [\apj]
  {10.1088/0004-637X/787/2/126}, \href
  {https://ui.adsabs.harvard.edu/abs/2014ApJ...787..126L} {787, 126}

\bibitem[\protect\citeauthoryear{{MacDonald} \& {Gizis}}{{MacDonald} \&
  {Gizis}}{2018}]{macd18}
{MacDonald} J.,  {Gizis} J.,  2018, \mn@doi [\mnras] {10.1093/mnras/sty1888},
  \href {http://adsabs.harvard.edu/abs/2018MNRAS.480.1711M} {480, 1711}

\bibitem[\protect\citeauthoryear{{Mace} et~al.,}{{Mace} et~al.}{2013}]{mace13}
{Mace} G.~N.,  et~al., 2013, \mn@doi [\apj] {10.1088/0004-637X/777/1/36}, \href
  {https://ui.adsabs.harvard.edu/abs/2013ApJ...777...36M} {777, 36}

\bibitem[\protect\citeauthoryear{{McMahon}, {Banerji}, {Gonzalez}, {Koposov},
  {Bejar}, {Lodieu}, {Rebolo}  \& {VHS Collaboration}}{{McMahon}
  et~al.}{2013}]{mcma13}
{McMahon} R.~G.,  {Banerji} M.,  {Gonzalez} E.,  {Koposov} S.~E.,  {Bejar}
  V.~J.,  {Lodieu} N.,  {Rebolo} R.,   {VHS Collaboration} 2013, The Messenger,
  \href {http://adsabs.harvard.edu/abs/2013Msngr.154...35M} {154, 35}

\bibitem[\protect\citeauthoryear{{Monet}, {Dahn}, {Vrba}, {Harris}, {Pier},
  {Luginbuhl}  \& {Ables}}{{Monet} et~al.}{1992}]{mone92}
{Monet} D.~G.,  {Dahn} C.~C.,  {Vrba} F.~J.,  {Harris} H.~C.,  {Pier} J.~R.,
  {Luginbuhl} C.~B.,   {Ables} H.~D.,  1992, \mn@doi [\aj] {10.1086/116091},
  \href {http://adsabs.harvard.edu/abs/1992AJ....103..638M} {103, 638}

\bibitem[\protect\citeauthoryear{{Newton}, {Charbonneau}, {Irwin}  \&
  {Mann}}{{Newton} et~al.}{2015}]{newt15}
{Newton} E.~R.,  {Charbonneau} D.,  {Irwin} J.,   {Mann} A.~W.,  2015, \mn@doi
  [\apj] {10.1088/0004-637X/800/2/85}, \href
  {https://ui.adsabs.harvard.edu/abs/2015ApJ...800...85N} {800, 85}

\bibitem[\protect\citeauthoryear{{Pavlenko}, {Zhang}, {G{\'a}lvez-Ortiz},
  {Kushniruk}  \& {Jones}}{{Pavlenko} et~al.}{2015}]{pavl15}
{Pavlenko} Y.~V.,  {Zhang} Z.~H.,  {G{\'a}lvez-Ortiz} M.~C.,  {Kushniruk}
  I.~O.,   {Jones} H.~R.~A.,  2015, \mn@doi [\aap]
  {10.1051/0004-6361/201526810}, \href
  {http://adsabs.harvard.edu/abs/2015A%26A...582A..92P} {582, A92}

\bibitem[\protect\citeauthoryear{{Pinfield} et~al.,}{{Pinfield}
  et~al.}{2012}]{pinf12}
{Pinfield} D.~J.,  et~al., 2012, \mn@doi [\mnras]
  {10.1111/j.1365-2966.2012.20549.x}, \href
  {https://ui.adsabs.harvard.edu/abs/2012MNRAS.422.1922P} {422, 1922}

\bibitem[\protect\citeauthoryear{{Rebassa-Mansergas}, {Parsons}, {Dhillon},
  {Ren}, {Littlefair}, {Marsh}  \& {Torres}}{{Rebassa-Mansergas}
  et~al.}{2019}]{reba19}
{Rebassa-Mansergas} A.,  {Parsons} S.~G.,  {Dhillon} V.~S.,  {Ren} J.,
  {Littlefair} S.~P.,  {Marsh} T.~R.,   {Torres} S.,  2019, \mn@doi [Nature
  Astronomy] {10.1038/s41550-019-0746-7}, \href
  {https://ui.adsabs.harvard.edu/abs/2019NatAs...3..553R} {3, 553}

\bibitem[\protect\citeauthoryear{{Reid} \& {Hawley}}{{Reid} \&
  {Hawley}}{2005}]{reid05}
{Reid} I.~N.,  {Hawley} S.~L.,  2005, {New light on dark stars : red dwarfs,
  low-mass stars, brown dwarfs}.
Springer-Verlag Berlin Heidelberg, \mn@doi{10.1007/3-540-27610-6}

\bibitem[\protect\citeauthoryear{{Reid}, {Hawley}  \& {Gizis}}{{Reid}
  et~al.}{1995}]{reid95}
{Reid} I.~N.,  {Hawley} S.~L.,   {Gizis} J.~E.,  1995, \mn@doi [\aj]
  {10.1086/117655}, \href {http://adsabs.harvard.edu/abs/1995AJ....110.1838R}
  {110, 1838}

\bibitem[\protect\citeauthoryear{{Rojas-Ayala}, {Covey}, {Muirhead}  \&
  {Lloyd}}{{Rojas-Ayala} et~al.}{2012}]{roja12}
{Rojas-Ayala} B.,  {Covey} K.~R.,  {Muirhead} P.~S.,   {Lloyd} J.~P.,  2012,
  \mn@doi [\apj] {10.1088/0004-637X/748/2/93}, \href
  {https://ui.adsabs.harvard.edu/abs/2012ApJ...748...93R} {748, 93}

\bibitem[\protect\citeauthoryear{{Ruiz}}{{Ruiz}}{1996}]{ruiz96}
{Ruiz} M.~T.,  1996, \mn@doi [\aj] {10.1086/117871}, \href
  {http://adsabs.harvard.edu/abs/1996AJ....111.1267R} {111, 1267}

\bibitem[\protect\citeauthoryear{{Russell}}{{Russell}}{1914}]{russ14}
{Russell} H.~N.,  1914, Popular Astronomy, \href
  {https://ui.adsabs.harvard.edu/abs/1914PA.....22..275R} {22, 275}

\bibitem[\protect\citeauthoryear{{Savcheva}, {West}  \& {Bochanski}}{{Savcheva}
  et~al.}{2014}]{savc14}
{Savcheva} A.~S.,  {West} A.~A.,   {Bochanski} J.~J.,  2014, \mn@doi [\apj]
  {10.1088/0004-637X/794/2/145}, \href
  {http://adsabs.harvard.edu/abs/2014ApJ...794..145S} {794, 145}

\bibitem[\protect\citeauthoryear{{Scholz}, {Lehmann}, {Matute}  \&
  {Zinnecker}}{{Scholz} et~al.}{2004}]{scho04}
{Scholz} R.-D.,  {Lehmann} I.,  {Matute} I.,   {Zinnecker} H.,  2004, \mn@doi
  [\aap] {10.1051/0004-6361:20041059}, \href
  {http://adsabs.harvard.edu/abs/2004A%26A...425..519S} {425, 519}

\bibitem[\protect\citeauthoryear{{Schweitzer}, {Scholz}, {Stauffer}, {Irwin}
  \& {McCaughrean}}{{Schweitzer} et~al.}{1999}]{schw99}
{Schweitzer} A.,  {Scholz} R.-D.,  {Stauffer} J.,  {Irwin} M.,   {McCaughrean}
  M.~J.,  1999, \aap, \href
  {http://adsabs.harvard.edu/abs/1999A%26A...350L..62S} {350, L62}

\bibitem[\protect\citeauthoryear{{Sivarani}, {L{\'e}pine}, {Kembhavi}  \&
  {Gupchup}}{{Sivarani} et~al.}{2009}]{siva09}
{Sivarani} T.,  {L{\'e}pine} S.,  {Kembhavi} A.~K.,   {Gupchup} J.,  2009,
  \mn@doi [\apjl] {10.1088/0004-637X/694/2/L140}, \href
  {http://adsabs.harvard.edu/abs/2009ApJ...694L.140S} {694, L140}

\bibitem[\protect\citeauthoryear{{Vernet} et~al.,}{{Vernet}
  et~al.}{2011}]{vern11}
{Vernet} J.,  et~al., 2011, \mn@doi [\aap] {10.1051/0004-6361/201117752}, \href
  {http://adsabs.harvard.edu/abs/2011A%26A...536A.105V} {536, A105}

\bibitem[\protect\citeauthoryear{{Woolf} \& {Wallerstein}}{{Woolf} \&
  {Wallerstein}}{2005}]{wool05}
{Woolf} V.~M.,  {Wallerstein} G.,  2005, \mn@doi [\mnras]
  {10.1111/j.1365-2966.2004.08515.x}, \href
  {https://ui.adsabs.harvard.edu/abs/2005MNRAS.356..963W} {356, 963}

\bibitem[\protect\citeauthoryear{{Woolf} \& {Wallerstein}}{{Woolf} \&
  {Wallerstein}}{2006}]{wool06}
{Woolf} V.~M.,  {Wallerstein} G.,  2006, \mn@doi [\pasp] {10.1086/498459},
  \href {http://adsabs.harvard.edu/abs/2006PASP..118..218W} {118, 218}

\bibitem[\protect\citeauthoryear{{Woolf}, {L{\'e}pine}  \&
  {Wallerstein}}{{Woolf} et~al.}{2009}]{wool09}
{Woolf} V.~M.,  {L{\'e}pine} S.,   {Wallerstein} G.,  2009, \mn@doi [\pasp]
  {10.1086/597433}, \href {http://adsabs.harvard.edu/abs/2009PASP..121..117W}
  {121, 117}

\bibitem[\protect\citeauthoryear{{York} et~al.,}{{York} et~al.}{2000}]{york00}
{York} D.~G.,  et~al., 2000, \mn@doi [\aj] {10.1086/301513}, \href
  {https://ui.adsabs.harvard.edu/abs/2000AJ....120.1579Y} {120, 1579}

\bibitem[\protect\citeauthoryear{{Zhang}}{{Zhang}}{2018}]{zha18c}
{Zhang} Z.,  2018, in 20th Cambridge Workshop on Cool Stars, Stellar Systems
  and the Sun. p.~44 (\mn@eprint {arXiv} {1810.07071}),
  \mn@doi{10.5281/zenodo.1463229}

\bibitem[\protect\citeauthoryear{{Zhang} et~al.,}{{Zhang} et~al.}{2010}]{zha10}
{Zhang} Z.~H.,  et~al., 2010, \mn@doi [\mnras]
  {10.1111/j.1365-2966.2010.16394.x}, \href
  {http://adsabs.harvard.edu/abs/2010MNRAS.404.1817Z} {404, 1817}

\bibitem[\protect\citeauthoryear{{Zhang} et~al.,}{{Zhang} et~al.}{2013}]{zha13}
{Zhang} Z.~H.,  et~al., 2013, \mn@doi [\mnras] {10.1093/mnras/stt1030}, \href
  {http://adsabs.harvard.edu/abs/2013MNRAS.434.1005Z} {434, 1005}

\bibitem[\protect\citeauthoryear{{Zhang} et~al.,}{{Zhang}
  et~al.}{2017a}]{prime1}
{Zhang} Z.~H.,  et~al., 2017a, \mn@doi [\mnras] {10.1093/mnras/stw2438}, \href
  {http://adsabs.harvard.edu/abs/2017MNRAS.464.3040Z} {464, 3040}

\bibitem[\protect\citeauthoryear{{Zhang}, {Homeier}, {Pinfield}, {Lodieu},
  {Jones}, {Allard}  \& {Pavlenko}}{{Zhang} et~al.}{2017b}]{prime2}
{Zhang} Z.~H.,  {Homeier} D.,  {Pinfield} D.~J.,  {Lodieu} N.,  {Jones}
  H.~R.~A.,  {Allard} F.,   {Pavlenko} Y.~V.,  2017b, \mn@doi [\mnras]
  {10.1093/mnras/stx350}, \href
  {http://adsabs.harvard.edu/abs/2017MNRAS.468..261Z} {468, 261}

\bibitem[\protect\citeauthoryear{{Zhang} et~al.,}{{Zhang}
  et~al.}{2018a}]{prime3}
{Zhang} Z.~H.,  et~al., 2018a, \mn@doi [\mnras] {10.1093/mnras/sty1352}, \href
  {http://adsabs.harvard.edu/abs/2018MNRAS.479.1383Z} {479, 1383}

\bibitem[\protect\citeauthoryear{{Zhang} et~al.,}{{Zhang}
  et~al.}{2018b}]{prime4}
{Zhang} Z.~H.,  et~al., 2018b, \mn@doi [\mnras] {10.1093/mnras/sty2054}, \href
  {http://adsabs.harvard.edu/abs/2018MNRAS.480.5447Z} {480, 5447}

\bibitem[\protect\citeauthoryear{{Zhang} et~al.,}{{Zhang}
  et~al.}{2019a}]{zhan19}
{Zhang} S.,  et~al., 2019a, \mn@doi [\apjs] {10.3847/1538-4365/aafb32}, \href
  {http://adsabs.harvard.edu/abs/2019ApJS..240...31Z} {240, 31}

\bibitem[\protect\citeauthoryear{{Zhang}, {Burgasser}, {G{\'a}lvez-Ortiz},
  {Lodieu}, {Zapatero Osorio}, {Pinfield}  \& {Allard}}{{Zhang}
  et~al.}{2019b}]{prime6}
{Zhang} Z.~H.,  {Burgasser} A.~J.,  {G{\'a}lvez-Ortiz} M.~C.,  {Lodieu} N.,
  {Zapatero Osorio} M.~R.,  {Pinfield} D.~J.,   {Allard} F.,  2019b, \mn@doi
  [\mnras] {10.1093/mnras/stz777}, \href
  {http://adsabs.harvard.edu/abs/2019MNRAS.486.1260Z} {486, 1260}

\bibitem[\protect\citeauthoryear{{Zhang}, {Burgasser}  \& {Smith}}{{Zhang}
  et~al.}{2019c}]{prime5}
{Zhang} Z.~H.,  {Burgasser} A.~J.,   {Smith} L.~C.,  2019c, \mn@doi [\mnras]
  {10.1093/mnras/stz659}, \href
  {http://adsabs.harvard.edu/abs/2019MNRAS.486.1840Z} {486, 1840}

\bibitem[\protect\citeauthoryear{{Zhong} et~al.,}{{Zhong}
  et~al.}{2015}]{zhon15}
{Zhong} J.,  et~al., 2015, \mn@doi [\aj] {10.1088/0004-6256/150/2/42}, \href
  {http://adsabs.harvard.edu/abs/2015AJ....150...42Z} {150, 42}

\makeatother
\end{thebibliography}

\label{lastpage}
\end{document}